\documentclass[showpacs,prb,letterpaper,floatfix,nobalancelastpage,twocolumn]{revtex4}
\usepackage{dcolumn}
\usepackage{bm}
\usepackage{amsmath}
\usepackage{bm,braket}
\usepackage{color}
\usepackage{float}
\usepackage{graphicx,subfigure}
\usepackage{amssymb}

\begin{document}


\title{Influence of electron-acoustic phonon scattering on intensity
power broadening in a coherently driven quantum-dot cavity system}
\author{C. Roy}
\email{chiranjeeb.roy@queensu.ca}
\author{S. Hughes}
\email{shughes@physics.queensu.ca}
\affiliation{Department of Physics, Engineering Physics and Astronomy,
Queen's University, Kingston, Ontario, Canada K7L 3N6}

\begin{abstract}

We present a   quantum optics formalism to study intensity power broadening of a semiconductor quantum dot interacting with an acoustic phonon bath and a high $Q$ microcavity. Power broadening is investigated using a time-convolutionless master equation in the polaron frame which allows for a nonperturbative treatment of the interaction of the quantum dot with the phonon reservoir. We calculate the full non-Lorentzian photoluminescence (PL) lineshapes and numerically extract the intensity linewidths of the quantum dot exciton and the cavity mode as a function of pump rate and temperature. For increasing field strengths, multiphonon and multiphoton effects
are found to be important, even for phonon bath temperatures as low as 4~K. We show  that the interaction of the quantum dot with the phonon reservoir introduces
pronounced features in the power broadened PL lineshape, enabling one to observe  clear signatures of  electron-phonon scattering. The  PL lineshapes from cavity pumping and  exciton pumping are found to be distinctly different, primarily since the latter is excited through the exciton-phonon reservoir. To help explain the underlying physics of phonon scattering on the power broadened lineshape, an effective phonon Lindblad master equation derived from the full time-convolutionless master equation is introduced; we  identify and calculate distinct Lindblad scattering contributions from electron-phonon interactions, including effects such as excitation-induced dephasing, incoherent exciton excitation and exciton-cavity feeding. Our effective phonon master equation is shown to reproduce the full intensity PL and the phonon-coupling effects very well,  suggesting that its general Lindblad form may find widespread use in semiconductor cavity-QED.

\end{abstract}


\maketitle

\section{Introduction}

Semiconductor quantum dots (QDs) embedded in microcavities have established themselves as a new paradigm in cavity  quantum electrodynamics (cavity-QED). Technological
progress in the design and fabrication of semiconductor cavity-QED systems has enabled them to be used as components 
in quantum information processing~\cite{hohenester,entangled1} and for the generation of  {\em indistinguishable photons}~\cite{entangled2, entangled3, ates1, santori}. These quantum applications
 require robust cavity-QED based QD devices that rest on the ability to manipulate and control the underlying quantum processes. Such quantum control is usually obtained when the cavity and QD are in the intermediate to strong coupling regime~\cite{SC:StrongCoupling1,SC:StrongCoupling2,press}. Recent experimental studies have  focussed on resonance fluorescence of a QD coupled to a cavity mode~\cite{muller,flagg,vamivakas}, and significant progress has  been made in the study of an off-resonant QD cavity system that is used to observe resonance fluorescence of a single photon emitter~\cite{ates2, jelena_arka}. Semiconductor micropillar systems are particularly attractive since a geometrical separation between the pump field and emitted fluorescence signal can be  made, facilitating  nonlinear quantum optical studies such as intensity power broadening~\cite{ates2}.

For semiconductor cavity-QED systems, signatures of acoustic phonon processes have been noted with {\em incoherent} excitation, resulting in off-resonant ``cavity feeding''~\cite{HennessyNature:2007,KaniberPRB:2008,RuthOE:2007,SufczynskiPRL:2009,ota,TawaraOE:2009,Dalacu:PRB2010,Calic:PRL11} and an asymmetric ({\em on-resonance}) vacuum Rabi doublet~\cite{hughes1,ota,hughes2}. Various phonon-coupling models have been developed to try and explain these features~\cite{JiaoJPC2008,ota,HohenesterPRB:2009,HohenesterPRB:2010,kaer,hughes2,SavonaPRB:2010};
for example,
data obtained for the linear spectrum of single site-selected dots in cavities show good agreement with
photon Green function theories---where the phonon coupling is included
as a  self-energy correction to the spectrum~\cite{hughes1,hughes2,Calic:PRL11}.
Recently, several works have also experimentally
investigated {\em coherent} power (intensity) broadening in semiconductor cavity-QED systems.
For example, Majumdar {\em et. al.}~\cite{jelena1} studied the role of phonon-mediated dot-cavity coupling
on the power broadened PL intensity for a planar photonic crystal system; with experiments  performed at
temperatures of $30$-$55$~K on self-assembled InAs QDs,
the cavity-emitted intensity PL were found to have extraneously broadened linewidths
relative to a bare QD (when compared with calculations from a  simple atomic ME).
While {\em additional} coupling may occur, e.g., from the
QD to the continuum states due to the presence of the nearby wetting layer (if it exists),
or due to Auger scattering~\cite{auger, auger1, auger2, auger3}, these processes are usually more important
for incoherent excitation.
 For near-resonant {\em coherent}  excitation,
Ulhaq {\em et. al.}~\cite{ates3} have demonstrated that
dephasing and coupling due to acoustic phonons is likely the primary
(and {\em intrinsic}) mechanism that couples the QD and the spectrally detuned cavity mode; their experiments were performed using self-assembled InGaAs/GaAs QDs 
embedded in a single
cavity  layer
 of a semiconductor micropillar cavity.
While the important  role of electron-phonon scattering on the linear absorption and emission spectra of
self-assembled QDs is now becoming better established~\cite{besombes,Favero:PRB03,Peter:PRB04}, there appears to be little
theoretical work describing phonon effects on PL power broadening in a
semiconductor cavity-QED system.

Nonlinear resonance fluorescence of an InGaAs QD embedded in a high-quality micropillar cavity was recently investigated by Ulrich {\em et al.}~\cite{stuttgart_prl}, where, in contrast to atomic cavity-QED, a clear indication of excitation-induced dephasing (EID) was found to manifest in a Mollow triplet spectra with pump-induced spectral sideband broadening.
Without cavity interactions, it is well known that the interaction of the driven QD with the underlying phonon reservoir 
can introduce additional dephasing processes and acoustic phonon sidebands~\cite{besombes,Favero:PRB03,Peter:PRB04}. In a cavity system, these phonon processes can also result in significant coupling
 between a non-resonant cavity and a QD exciton.  Very recently, a polaron  master equation (ME) description of 
 phonon-induced EID in QDs and cavity-QED  was described
 by Roy and Hughes~\cite{roy_hughes}; McCutcheon and Nazir 
 have also adopted a polaron ME approach to describe pule-excited excitons (without
cavity interactions)~\cite{nazir2}. In light of these
phonon scattering studies and the emerging class of 
 semiconductor cavity-QED experiments,
 the inclusion of phonon-scattering in the theoretical description of PL power broadening in a cavity-QD system  is highly desired. More generally, one desires accurate quantum optical descriptions of the semiconductor cavity-QED system, where  important electron-phonon interactions are accounted for.

 In this paper,  we present a
 quantum  ME formalism to study the intensity power broadening of a
  semiconductor cavity-QED system and identify the qualitative features of power broadening in the intensity PL introduced due to 
  electron-phonon interactions.
 We exploit a time-convolutionless ME (i.e., local in time) for the reduced density matrix  of the dot-cavity subsystem, where the system-bath incoherent interaction is treated to second order~\cite{roy_hughes, nazir2}. The perturbative treatment is  performed in the polaron
 frame which allows us to study the effects of phonon dephasing on the coherent part of the Hamiltonian
 exactly.
 Importantly, the cw laser driving the QD introduces additional EID effects in addition to pure dephasing due to the phonon reservoir~\cite{roy_hughes, nazir1}.
In the appropriate limits, the model fully recovers the independent boson model (IBM)~\cite{mahan,imamoglu,krum} and the Jaynes-Cummings model.
 The polaron transform is
  particularly convenient for studying QD cavity-QED systems as it eliminates the exciton-phonon coupling and introduces  a modified dot-cavity coupling and a modified radiative decay rate~\cite{mahan,imamoglu}; in addition, there is a phonon-induced renormalization of the QD resonance frequency through the polaron shift. In the case of an exciton driven system, the Rabi frequency of the cw laser is also renormalized by a temperature-dependent factor which essentially accounts for the dephasing of the cw drive due to phonon coupling.
 A similar polaron ME approach
 was previously derived
 by Wilson-Rae and Imamogl{\u u}~\cite{imamoglu}, who studied the linear absorption spectrum of a cavity-QED system; 
 however, their ME form~\cite{wurger} was non-local in time and is
 substantially more difficult to solve 
 than the time-convolutionless form~\cite{note3}.

  For our coherently-pumped  cavity-QED
  investigations, we consider two distinctly different pumping scenarios: $(i)$ the QD is
   driven by a coherent continuous wave (cw) laser field, and $(ii)$ the cavity mode 
is driven by a coherent 
   cw laser field.
   We describe the generic features arising due to the relative interplay between phonon-induced
   dot cavity coupling and EID in the case of a QD driven system, and
 compare and contrast with power broadening for a cavity driven system; we also  discuss the
 differences
 in the integrated PL (IPL) for a dot-driven and cavity-driven system.
 For strong coherent drives (fields), the intensity PL contain  significant phonon bath signatures over a wide range of frequencies.
 To help explain the effects
 of phonon scattering in these systems, we  also derive an effective phonon ME, of the Lindblad form
 which is shown, in certain regimes, to yield very good agreement with
 the full polaron ME.

Our paper is organized as follows. In Sec.~\ref{model} we present the model Hamiltonian and derive a time-convolutionless polaron ME where electron-phonon interactions are included to all orders. In Sec.~\ref{Lindblad} we introduce an {\em effective} phonon-modified Lindblad ME and compare it to the full time-convolutionless solution; the effective Lindblad ME is shown to yield good agreement with the
time-convolutionless ME, and we use it to describe the various 
phonon scattering processes.
 In Sec.~\ref{results}(A-E) 
  we present and discuss our numerical results of the power broadening lineshape for both QD-driven and cavity-driven systems. In Sec.~\ref{conclusions}  we present our conclusions.
Appendix~\ref{appendix1} provides some technical details
about the derivation of our effective 
phonon scattering rates and Lindblad ME.

\section{General Theory and Polaron Master Equation Model}
\label{model}

\begin{figure}[t]
\centering\includegraphics[width=0.99\columnwidth]{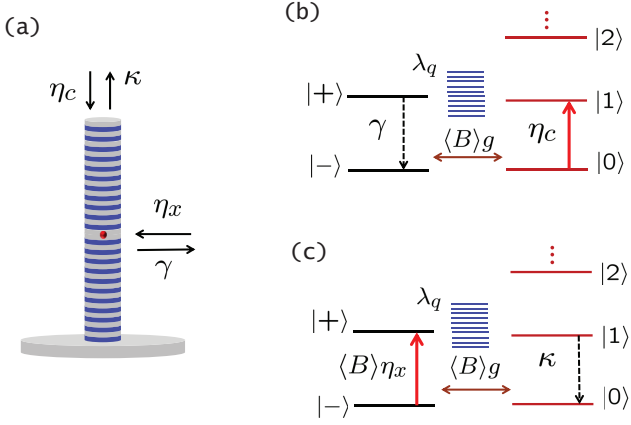}
\caption{(a) (Color online) Schematic of an example semiconductor cavity used in cavity-QED
(micropillar system), containing a coupled QD and driven by a cw laser ($\eta_x$, from the side,
or $\eta_c$, from the top). The micropillar system is advantageous to study PL power broadening
since, e.g., exciton broadening can be directly studied through cavity emission---since these  input/output channels are geometrically
 decoupled.
 (b) Schematic of a cavity driven cavity-QED system, where
 $\vert +\rangle$ denotes the excited QD state, $\vert -\rangle$ denotes the ground state, $\vert 1\rangle$ and $\vert 0\rangle$ 
  represent the first excited and ground state of the cavity mode. Also shown is the phonon reservoir as blue (multiple) lines. The
 terms  $\braket{B}\!g$ and $\braket{B}\!\eta_x$ represent a phonon-modified coherent reduction
 ($\braket{B}(T) \leq 1$) in the exciton-cavity coupling rate and the exciton pump rate, respectively (see text for details).
  (c) Schematic of a dot-driven cavity-QED system with same energy level notation as in (b). }
 \label{fig:1n}
\end{figure}

The dynamics of a strongly confined
QD can be modeled by considering a
 quantized electron-hole excitation,  where the electron occupies
a conduction band state and the hole occupies a valence band state.
Neglecting quantum spin, the dominant features of a strongly confined QD can be described by the two lowest energy bound states.
 This 
 two-level model is then conditioned by the interaction of the electrons with the lattice modes
  of vibration, i.e., the acoustic phonons. When the effective two-level system is driven by a cw laser field, power broadening 
  may be substantially modified by the coupling of the QD to the phonon modes~\cite{mogilevtsev1, ramsay2}.
 Figure~\ref{fig:1n} shows a schematic
 of a semiconductor cavity-QED system [Fig.~\ref{fig:1n}(a)], and an energy-level diagram associated
 with cavity-pumping [Fig.~\ref{fig:1n}(b)] and exciton pumping [Fig.~\ref{fig:1n}(c)]; the various parameters
 in the figures  will be introduced below. The semiconductor cavity system of interest
 could be a  micropillar cavity system [cf.~Fig.~\ref{fig:1n}(a)] which allows one to excite and measure through different
 photon reservoirs~\cite{stuttgart_prl} (e.g., cavity pumping and exciton emission).

Working in a frame rotating with respect to the laser pump frequency, $\omega_L$, we first introduce the model Hamiltonian describing a cavity-QED system
where the QD interacts with an acoustic phonon reservoir:
\begin{align}
\label{sec1eq1}
H&=\hbar\Delta_{xL}\hat{\sigma}^{+}\hat{\sigma}^{-}+\hbar\Delta_{cL}\hat{a}^{\dagger}\hat{a} +
\hbar g(\hat{\sigma}^{+}\hat{a}+\hat{a}^{\dagger}\hat{\sigma}^{-}) \nonumber \\
&+H^{x/c}_{\rm{drive}} +\hat{\sigma}^{+}\hat{\sigma}^{-}\sum_{q}\hbar\lambda_{q}(\hat{b}_{q}
+\hat{b}_{q}^{\dagger})+\sum_{q}\hbar\omega_{q}\hat{b}_{q}^{\dagger}\hat{b}_{q}\, ,
\end{align}
where  $\hat{b}_{q}(\hat{b}_{q}^{\dagger})$ are the annihilation and creation operators of the phonon reservoir,
 $\hat a$ is the {\em leaky} cavity mode annihilation operator,  $\hat{\sigma}^+$ (annihilation) and $\hat{\sigma}^-$
 (creation)
 are the Pauli operators of
the electron-hole pair or exciton; $\Delta_{\alpha L}\equiv \omega_\alpha-\omega_L$ ($\alpha =x,c$)
are the detunings of the exciton ($\omega_{x}$) and cavity ($\omega_{c}$) from the coherent pump laser ($\omega_{L}$),  and $g$ is the cavity-exciton  coupling strength.
The  pump term, $H^{x/c}_{\rm{drive}}$,
 accounts for the {\em coherent} drive on the cavity-QED system; for a QD (exciton) driven system, $H^{x}_{\rm{drive}}=\hbar \eta_{x}(\hat{\sigma}^{+}+\hat{\sigma}^{-})$, while for a cavity  driven system,
 $H^{c}_{\rm{drive}}=\hbar \eta_{c}(\hat{a}+\hat{a}^{\dagger})$. The defined pump rate,
$\eta_{x/c}$, is two times the classical Rabi frequency.

Transforming to the polaron frame, we  eliminate the QD-phonon coupling and introduce a renormalized dot-cavity coupling
strength~\cite{kaer}. For the case of the QD driven system, the polaron transformation  also results in a renormalized Rabi
frequency, defined below. The polaron transformation~\cite{mahan,hohenester} can be written as
\begin{equation}
\label{sec2eq7}
H^{\prime}=\exp({S})H\exp(-{S}),
\end{equation}
where
\begin{equation}
\label{sec2eq8}
{S}=\hat{\sigma}^{+}\hat{\sigma}^{-}\sum_{q}
\frac{\lambda_{q}}{\omega_{q}}(\hat{b}_{q}^{\dagger}-\hat{b}_{q}).
\end{equation}
The transformed Hamiltonian becomes
\begin{subequations}
\begin{align}
\label{sec3eq1}
H^{\prime}_{sys}  & =    \hbar(\Delta_{xL}-\Delta_{P})\hat{\sigma}^{+}\hat{\sigma}^{-}
+\hbar\Delta_{cL} \hat{a}^{\dagger}\hat{a}+\langle B\rangle \hat{X}_{g}, \\
H^{\prime}_{bath} & =  \sum_{q}\hbar\omega_{q}\hat{b}_{q}^{\dagger}\hat{b}_{q},\\
H^{\prime}_{int}  & =  \hat{X}_{g}\hat{\zeta}_{g}+\hat{X}_{u}\hat{\zeta}_{u},
\end{align}
\end{subequations}
with
\begin{subequations}
\begin{align}
\label{sec3eq2a}
\hat{B}_{\pm}&=\exp\left (\pm\sum_{q}\frac{\lambda_{q}}{\omega_{q}}(\hat{b}_{q}-\hat{b}_{q}^{\dagger}) \right),
\\
\hat{\zeta}_{g}&=\frac{1}{2}(\hat{B}_{+}+\hat{B}_{-}-2\langle B\rangle),  \\
\hat{\zeta}_{u}&=\frac{1}{2i}(\hat{B}_{+}-\hat{B}_{-}).
\end{align}
\end{subequations}
The polaron shift,
\begin{align}
\Delta_P=\int^{\infty}_{0}d\omega\frac{J(\omega)}{\omega},
\end{align}
and 
the thermally-averaged bath displacement operator~\cite{mahan},
\begin{subequations}
\begin{align}
\label{sec3eq5a}
\braket{B}&=\exp\left [ -\frac{1}{2}\int^{\infty}_{0}d\omega\frac{J(\omega)}{\omega^{2}}\coth(\beta\hbar\omega/2) \right ],\\
&=
\exp \left [ - \frac{1}{2} \sum_q \left (\frac{\lambda_q}{\omega_q}\right )^2 ( 2\bar n_q+1) 
\right ],  \\
& =\langle {B}_{+}\rangle=\langle {B}_{-}\rangle,
\end{align}
\end{subequations}
where $\bar n_q \equiv \braket{\hat b_q^\dagger \hat b_q} = [e^{\beta \hbar \omega_q}-1]^{-1}$
is the mean phonon occupation number (Bose-Einstein distribution) at a bath temperature, $T=1/k_b \beta$.
For clarity, we will henceforth assume that the  polaron shift is implicitly included in
our definition of $\omega_{x}$ (one should, however, keep in mind that this shift is
temperature dependent).
For a dot (exciton) driven system, $\hat X_{g}$ and $\hat X_{u}$ are defined through
\begin{subequations}
\begin{align}
\label{sec3eq2b}
\hat{X}_{g}& = \hbar g(\hat{a}^{\dagger}\hat{\sigma}^{-}+\hat{\sigma}^{+}\hat{a})
+\hbar \eta_{x}(\hat{\sigma}^{-}+\hat{\sigma}^{+}),  \\
\hat{X}_{u}& = i\hbar[g(\hat{\sigma}^{+}\hat{a}-\hat{a}^{\dagger}\hat{\sigma}^{-})
+i\hbar \eta_{x}(\hat{\sigma}^{+}-\hat{\sigma}^{-})],
\end{align}
\end{subequations}
and for a cavity driven system,
\begin{subequations}
\begin{align}
\label{sec3eq2b1}
\hat{X}_{g}& = \hbar g(\hat{a}^{\dagger}\hat{\sigma}^{-}+\hat{\sigma}^{+}\hat{a})  , \\
\hat{X}_{u}& = i\hbar g(\hat{\sigma}^{+}\hat{a}-\hat{a}^{\dagger}\hat{\sigma}^{-}).
\end{align}
\end{subequations}

Worth to note is the slightly unusual definition of the system Hamiltonian, Eq.~(\ref{sec3eq1}).
The {usual} (but in general, {\em incorrect}) decomposition of the system Hamiltonian to include
only the noninteracting QD and cavity parts does not take into account the effect of the
coherent cw drive on the system Hamiltonian. As the cavity and the QD systems are internally coupled,
 as discussed by Carmichael and Walls~\cite{carmichael}, this leads to violation of
  {\em detailed balance}. The system Hamiltonian written above leads to the correct form of
  the density operator while preserving detailed balance. Moreover, it  includes the effect
  of dot-cavity coupling and the dot-cw driving on the coherent part of the Hamiltonian to all orders.

Next, we unitarily transform to a frame of reference defined by this system Hamiltonian 
which we will use to obtain a
time-convolutionless ME; the net effect of this transform, e.g., in the case of resonance fluorescence through an exciton driven system,
 results in Mollow triplet peaks that
 sample the asymmetric phonon bath at the dressed eigenfrequencies---as determined by the modified system Hamiltonian.
 Somewhat similar techniques (i.e., bath sampling at the dressed resonances) have been employed to study atomic dynamics in
generalized ({\em engineered}) photon reservoirs, including
photonic band gap materials~\cite{sajeev1} and squeezed reservoirs~\cite{tanas1}.

Phenomenologically,
we 
include the radiative decay of the QD and the cavity mode as Liouvillian superoperators
acting on the reduced system density matrix~\cite{ota}. In addition,  we incorporate
an additional pure dephasing process beyond the IBM with a rate $\gamma^{\prime}$---this accounts for the
broadening of the zero-phonon line (ZPL) with increasing temperatures~\cite{besombes,BorriPRL:2001,Rudin:PRL06,Zimmermann:PRL04,zpl1,10,11,12,13,14}.
Though there is some controversy about what causes the broadening of the ZPL,
e.g., spectral diffusion, anharmonicity effects,\cite{10, 11} phonon scattering from interfaces,\cite{12, 13} and a modified phonon spectrum \cite{14}, it is well known that the ZPL broadens as a function of temperature with
a Lorentzian scattering process; thus, we treat the 
 broadening of the ZPL phenomenologically, while accounting
for broadening as a function of temperature similar to experiments~\cite{BorriPRL:2001,zpl1,ota},
with $\gamma'(T)$ scaling as $ \sim 1\mu eV/K$.
The various superoperators
act on the reduced system density matrix , and are defined
through
\begin{eqnarray}
\label{sec2eq34}
L(\rho)&=&\frac{\tilde{\gamma}}{2}(2\hat{\sigma}^{-}\rho\hat{\sigma}^{+}
-\hat{\sigma}^{+}\hat{\sigma}^{-}\rho-\rho\hat{\sigma}^{+}\hat{\sigma}^{-}) \nonumber \\
&+&\kappa(2\hat{a}\rho\hat{a}^{\dagger}-\hat{a}^{\dagger}\hat{a}\rho-\rho\hat{a}^{\dagger}\hat{a}) \nonumber \\
&+&
\frac{{\gamma'}}{2}(2\hat{\sigma}_{11}\rho\hat{\sigma}_{11}
-\hat{\sigma}_{11}\hat{\sigma}_{11}\rho-\rho\hat{\sigma}_{11}\hat{\sigma}_{11})\, ,
\end{eqnarray}
where $2\kappa$ is the cavity decay rate, $\tilde{\gamma}=\gamma\langle B\rangle^{2}$
is the radiative decay rate, and $\hat \sigma_{11}=\hat \sigma^+\hat \sigma^-$. The radiative decay rate has an additional renormalization by a factor of $\langle B\rangle^{2}$,
which reduces the effective radiative decay rate in the presence of phonons~\cite{roy} .

We
then derive a time convolutionless ME for the reduced density operator, $\rho(t)$, of the cavity-QED system~\cite{nazir2}
in the second-order Born approximation (for incoherent bath coupling). The time-convolutionless form of the ME, though local in time,
is known to capture non-Markov effects due to the reservoir~\cite{breuer}. However, for our analysis,
 we will  make a Markov approximation as typical phonon processes are substantially faster
 (i.e., a few ps) than the relevant system dynamics by at least an order of magnitude. This allows us to
 obtain effective rates which naturally depend on the spectral
 densities of the phonon spectral function that are locally sampled by the
 {\em dressed} resonances.
 We have  checked that
the Markov limit of the fully non-Markovian time-convolutionless ME
is rigorously valid for the system and excitation (cw) cases of interest. Thus, while it is straightforward
to carry out non-Markov calculations, it is  not necessary here---they give identical results.

In the interaction picture described by
$H^{\prime}_{sys}$, we consider the exciton-photon-phonon coupling $H_{int}^{\prime}$ to second order (Born approximation), and
trace over the phonon degrees of freedom to obtain a Markovian time convolutionless ME~\cite{nazir2,roy_hughes}:
\begin{eqnarray}
\label{sec3eq3}
\frac{\partial \rho}{\partial t}&=&\frac{1}{i\hbar}[H_{sys}^{\prime},\rho(t)]+L(\rho)-\frac{1}{\hbar^{2}}\int^{\infty}_{0}d\tau\sum_{m=g,u}
\bigg (G_{m}(\tau) \nonumber \\
&\times&\left [\hat{X}_{m},e^{-iH_{sys}^{\prime}\tau/\hbar}\hat{X}_{m}e^{iH_{sys}^{\prime}\tau/\hbar}\rho(t)\right]
+H.c. \bigg),
\end{eqnarray}
where $G_{g/u}(t)=\langle\zeta_{g/u}(t)\zeta_{g/u}(0)\rangle$.
The polaron Green functions are~\cite{mahan,imamoglu},
\begin{subequations}
\begin{align}
\label{sec3eq4}
G_{g}(t)&=\langle B\rangle^{2}\left (\cosh[\phi(t)]-1 \right ), \\
G_{u}(t)&=\langle B\rangle^{2}\sinh[\phi(t)],
\end{align}
\end{subequations}
which depend on the phonon correlation function,
\begin{subequations}
\begin{align}
\label{sec3eq5}
\phi(t)&=\int^{\infty}_{0}d\omega\frac{J(\omega)}{\omega^{2}}
\left [ \coth(\beta\hbar\omega/2)\cos(\omega t)-i\sin(\omega t)\right ],\\
&=\sum_q \left (\frac{\lambda_q}{\omega_q}\right )^2 \left [
(\bar n_q+1)e^{-i\omega_q t} + \bar n_q e^{i\omega_q t}
\right ],  
\end{align}
\end{subequations}
where $J(w)$ is the
 the characteristic phonon spectral function,
 defined in this work as
\begin{align}
\label{eq:J}
 J(\omega)=\alpha_{p}\,\omega^{3}\exp\left (-\frac{\omega^{2}}{2\omega_{b}^{2}} \right ).
 \end{align}
This  form of the spectral function [Eq.~(\ref{eq:J})]  describes the electron-LA(longitudinal acoustic)-phonon interaction via a deformation potential coupling
which is the
main source of dephasing in
self-assembled InAs/GaAs
QDs. For all our calculations that follow, we use parameters suitable for
  InAs/GaAs QDs~\cite{QDParams}, with
  $\omega_b=1~$meV ($\omega_{b}$ is a high frequency cutoff proportional to the inverse of the typical electronic localization length in the QD) and $\alpha_p/(2\pi)^2=0.06\,{\rm ps}^2$;
  these values vary somewhat in the literature, though
  we have taken ours from fitting recent experiments~\cite{hughes2,stuttgart_prl,roy_hughes}.
Using the parameters above, e.g.,
at $T=10~$K,
 yields a polaron shift, $\Delta_P\equiv 42~\mu eV$,
and a Franck-Condon renormalization, $\langle B\rangle=0.84$. With these phonon 
parameters, we already see that clearly the
coherent renormalization effects will be important for anlyzing
PL intensity for QD-cavity systems, even at relatively low
phonon bath temperatures.

We briefly mention that
there are other electron-acoustic phonon scattering models that can go beyond the
polaron ME approach.
 For example, McCutcheon {\em et. al.}~\cite{nazir3}
recently introduced a more general ME technique to describe the non-equilibrium dynamics of a QD system interacting with
 a phonon reservoir based on a variational formulation (with no
cavity coupling). This elegant approach extends the validity of the ME to parameter regimes, $\eta_x\ge\omega_{b}$, where
the ME in the polaron frame can break down. However, the pump parameter regimes that we study in this
work ($\eta_{x/c}\ll\omega_{b}$) are well within the domain of validity of our ME,
so we can safely use the polaron ME, while also accounting for cavity coupling~\cite{roy_hughes}.
A more general description of the system dynamics valid in all regimes can be obtained using a
quasi-adiabatic path integral approach~\cite{makri1, makri2}.
The benefits of our polaron ME is that the solution, even with  multiphonon
and multiphoton effects included,
is relatively straightforward,
and it has already been used  to
 help explain experiments for coherently-excited
dots in the regime of cavity-QED~\cite{stuttgart_prl,roy_hughes}. Moreover, as we will show
below, one can derive a user-friendly Lindblad ME that contains many of the key features
of phonon interactions in 
cavity-QED systems.

For numerical calculations, we solve the above ME with
 steady-state pumping (i.e., $\eta_{x/c}$ are time-independent), with the
exciton initially in the ground state.
Prior to these dynamical calculations,
we compute the phonon scattering terms
in Eq.~(\ref{sec3eq3}), whose solution
is naturally problem-dependent (through $H'_{sys}$).
Thus 
{there are no fixed phonon scattering rates
for analyzing QD power broadening as a function of pump power, as the phonon
scattering rates
are pump-dependent}. The same arguments apply for studying power broadening as a function of temperature; one must obtain the phonon-induced scattering rates for each pump value and temperature.
Experimentally, 
the  intensity PL lineshape is usually obtained by measuring
the QD exciton intensity ($I_{x}$) or cavity mode intensity ($I_c$) as a function of
increasing pump field.
To connect to these quantities, we
solve the above ME in a 
Jaynes-Cummings basis
with states $\ket{0},\ket{1L},\ket{1U},\ket{2L},\ket{2U},\cdots$,
and compute the steady-state
exciton and cavity photon populations, $\bar n_x \equiv \braket{\sigma^+\sigma^-}|_{ss} \propto I_x$
and $\bar n_c \equiv \braket{a^\dagger a}_{ss} \propto I_c$.
Defining the photon states $\ket{n}$, with $n=0,1,2, \cdots$,  and exciton states $\ket{+/-}$,
then the Jaynes-Cummings ladder states
are related to the bare states, e.g., through:
$\ket{0}=
\ket{-}\ket{0}$, $\ket{1L}=\frac{1}{\sqrt{2}}(\ket{-}\ket{1}-\ket{+}\ket{0})$,
$\ket{1U}=\frac{1}{\sqrt{2}}(\ket{-}\ket{1} + \ket{+}\ket{0})$.
%

In our calculations
we make partial use of the {\em quantum
optics toolbox} by Tan~\cite{QOToolbox}, and find
that truncation to two-photon-correlations (two photons or 5 states) is sufficient/necessary for all
the dot driven simulations, while truncation to
six-photon-correlations (six photons or 13 states) is sufficient/necessary for all
cavity calculations that follow. The role of multiphoton effects
depends on the value of the dot-cavity coupling rate
$g$, which we choose to be $g=20~\mu$eV---consistent with typical
semiconductor cavity-QED power broadening experiments (e.g., see Refs.~\onlinecite{jelena1, ates3}).
A  detailed discussion of the role of multiphoton,  and multiphonon processes, is presented
in Sec.~\ref{multiphoton_section}.


\section{\bf Effective phonon master equation of the Lindblad form}
\label{Lindblad}
Our polaron ME [Eq.~(\ref{sec3eq3})]  includes both coherent and incoherent contributions
from electron-phonon scattering, but some care and insight is needed in extracting the relevant
incoherent scattering rates.
It is therefore instructive to construct a simplified phonon-modified ME of the Lindblad form,
which we call an {\em effective} phonon master equation (EPME);
we do this by simplifying  the term, $e^{-iH_{sys}^{\prime}\tau/\hbar}\hat{X}_{m}e^{iH_{sys}^{\prime}\tau/\hbar}$,
appearing in
the full time-convolutionless ME [Eq.~(\ref{sec3eq3})].
The resulting Lindblad-form ME enables a very simple numerical solution
and facilitates the extraction of various
phonon-induced scattering rates in a clear and transparent way.
We expect that the integral in Eq.~(\ref{sec3eq3}) can be approximated,
 under certain circumstances, by only including the phase evolution of the operators $\hat{X}_{g,u}$ with
 respect to the noninteracting
part of the system evolution. Further, for a QD-driven system we only include terms proportional to
 $g^{2}$ and $\eta_{x}^{2}$ and
ignore cross terms proportional to $g\eta_{x}$; the inclusion of the cross terms do not preserve the Lindblad form
and contribute very little
to the overall broadening lineshape as can be demonstrated numerically. For a cavity-driven system, we again include
the phase evolution of
the operators $\hat{X}_{g,u}$ with respect to the noninteracting part of the system evolution; however, the effective
Lindblad description has
only contributions which are proportional to $g^2$---since $\hat{X}_{g,u}$ do not depend on $\eta_{c}$.
We will, of course, compare the EPME solution with the full  numerical solution of
the polaron time-convolutionless ME, i.e.,  Eq.~(\ref{sec3eq3}); the prime purpose of the  EPME is to help
elucidate the physics of phonon-induced incoherent scattering, though we will highlight regimes
where it can work quite well in accurately describing the 
full characteristics of the entire power-broadened PL lineshape.

\begin{figure}[t!]
\centering\includegraphics[width=1\columnwidth]{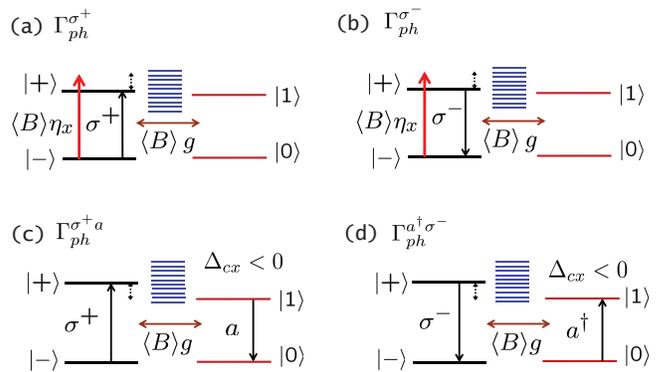}
\caption{(Color online) Schematic of the  phonon-scattering processes:
(a) $\Gamma_{ph}^{\sigma^{+}}$, (b) $\Gamma_{ph}^{\sigma^{-}}$,
(c) $\Gamma_{ph}^{\sigma^{+}a}$, and (d) $\Gamma_{ph}^{a^{\dagger}\sigma^{-}}$,
for a QD-driven system introduced in the
effective Lindblad description and defined in
Eqs.~(\ref{sec3eqfinal5})-(\ref{sec3eqfinal5R}).
The various symbols are the same as in Fig.~\ref{fig:1n}.
Processes (a) and (b) describe phonon-induced
incoherent excitation and excitation induced dephasing;
process (c) describes phonon-induced exciton excitation via cavity decay, though this process
has a small probability because of relatively large cavity decay rate ($\kappa$);
 process (d) describes phonon-induced cavity excitation through the decay of an exciton; at low
 temperature this (cavity {\em feeding}) process occurs primarily  via phonon {\em emission}, and will
 be more efficient for a cavity mode that is red shifted with respect to the
 exciton.
 The sign of the
 cavity detuning in (c-d) is just an example, and
 note that
 the phonon interactions with $\Delta_{cx}=\Delta$ or $\Delta_{cx}=-\Delta$ will be different (especially at lower
 temperatures).  }
\label{fig:2n}
\end{figure}

We postulate that the dynamics of the QD driven system can now be
approximately described through
\begin{equation}
\frac{\partial \rho}{\partial t}=\frac{1}{i\hbar}[H_{ sys}^{\rm eff},\rho(t)]+L(\rho)+L_{ ph}(\rho) ,
\label{effectiveME}
\end{equation}
where $L_{ ph}(\rho)$ (`$ph$' refers to phonon) is given by
\begin{eqnarray}
\label{sec3eqfinal3}
L_{ph}(\rho)&=&\frac{\Gamma_{ph}^{\sigma^{-}}}{2}L(\hat{\sigma}^{-})+\frac{\Gamma_{ph}^{\sigma^{+}}}{2}L(\hat{\sigma}^{+})
\nonumber \\
&+&\frac{\Gamma_{ph}^{\sigma^{+}a}}{2}L(\hat{\sigma}^{+}\hat{a})
+\frac{\Gamma_{ph}^{a^{\dagger}\sigma^{-}}}{2}L(\hat{a}^{\dagger}\hat{\sigma}^{-}),
\end{eqnarray}
and the superoperator $L(\hat{D})$ is defined as
\begin{equation}
\label{sec3eqfinal4}
L(\hat{D})=2\hat{D}\rho\hat{D^{\dagger}}-\hat{D^{\dagger}}\hat{D}\rho-\rho\hat{D^{\dagger}}\hat{D}.
\end{equation}
The above effective ME [Eq.~(\ref{effectiveME})] has a remarkably simple form, and its general format should be  familiar to many researchers
who have been using atomic cavity-QED models to connect to experimental
data using semiconductor  cavity-QED systems. However, it must be used with caution, as it 
is only valid within certain regimes where the above noted approximations are good. The phonon-mediated rates, which drive the effective Lindblad dynamics,
are derived to be (see Appendix~\ref{appendix1}):
\begin{align}
\label{sec3eqfinal5}
\Gamma_{ph}^{\sigma^{-}/\sigma^{+}}& =2\braket{B}^2\!\eta_{x}^{2}\,{\rm Re} \left [ \int_{0}^{\infty}d\tau\,
e^{\pm i\Delta_{xL}\tau}\! \left (e^{\phi(\tau)} -1 \right )\right], \\
\label{sec3eqfinal5R}
\Gamma_{ph}^{\sigma^{+}a/a^{\dagger}\sigma^{-}}& =2\braket{B}^2 \! g^{2}\,{\rm Re} \left [\int_{0}^{\infty}d\tau\,
e^{\pm i\Delta_{cx} \tau}\! \left (e^{\phi(\tau)}-1 \right )\right],
\end{align}
where $\Delta_{cx}=\omega_c-\omega_x$ is the cavity-exciton detuning. Figure \ref{fig:2n} shows a schematic of the various effective phonon-scattering processes: $\Gamma_{ph}^{\sigma^+}$ describes phonon-assisted {\em incoherent} excitation
and EID (pump-induced broadening);
$\Gamma_{ph}^{\sigma^-}$ described enhanced radiative decay and EID;
$\Gamma_{ph}^{\sigma^+a}$ describes (the somewhat unlikely scenario of) exciton excitation via the emission of a cavity photon, and $\Gamma_{ph}^{a^\dagger\sigma^-}$ describes the process of cavity excitation (cavity {\em feeding}) via the absorption of a photon. Importantly, all of these scattering events are driven by electron-phonon interactions
and they cause effects that are significantly different to simple pure dephasing models. In fact, pure
dephasing through $\gamma'$ only results in Lorentzian coupling, and is found to play  a minor role
in what follows below.

Our formalism above shows that for a cavity-driven system, $\Gamma_{ph}^{\sigma^{+}/\sigma^{-}}=0$ and $\Delta_{ph}^{\sigma^{-
}/\sigma^{+}}=0$, and there is no
phonon-induced EID due
 to the lack of any coupling of the drive with the phonon reservoir. We therefore expect (and find)

substantially different intensity power broadening between dot-driven and cavity-driven systems;  both
of these exciton-driven and cavity-driven models are also markedly different to simple atomic models.
We highlight that a similar exciton-cavity (feeding) rate, $\Gamma_{ph}^{a^\dagger\sigma^-}$, has been  derived by
Xue {\em et al.}~\cite{JiaoJPC2008} and by
Hohenester~\cite{HohenesterPRB:2010}, though these
were  obtained for an {\em undriven} cavity-QED system.
Both of these  useful approaches also
use a polaron frame to describe the incoherent scattering,
though the end equations have some
potential problems for small cavity-exciton detunings (where, admittedly, these effective
rates are at best approximate anyway), and neither approach
includes a coherent (temperature-dependent) reduction in $g\rightarrow \braket{B}\!g$.
 For example, Hohenester~\cite{HohenesterPRB:2010}
derives the following exciton-cavity feeding rate: 
$\Gamma_{ph}^{a^{\dagger}\sigma^{-}}=2  g^{2}\,{\rm Re}\!\left [\int_{0}^{\infty}d\tau\,
e^{-i\Delta_{cx} \tau}\! e^{-\phi(\tau)} \right], $
which has a similar form to our Eq.~(\ref{sec3eqfinal5R}) (apart from the sign of the phase and the need to subtract of a background
term for small detuning), but does not include
$\braket{B({\rm T})}^2$---an important 
 temperature-dependent term. The general predictions using this rate
 formula 
are consistent with 
experiments~\cite{HohenesterPRB:2009}, and
all these aforementioned polaron formalisms produce qualitatively the
same trend 
as a function of cavity-exciton detuning (compare
results in Refs.~\onlinecite{JiaoJPC2008,HohenesterPRB:2010} with those
in Figs.~\ref{fig:5n}-\ref{fig:6n}). 

An alternative effective Lindblad ME with
the same process identified in Refs.~\onlinecite{JiaoJPC2008,HohenesterPRB:2010}
and above (i.e., for exciton-cavity coupling), was recently presented by Majumdar {\em et. al.}\cite{jelena_arka} and used in part to study the role of phonon scattering for the cavity-emitted resonance fluorescence spectrum. The influence of phonons was included as two additional incoherent decay terms which were to second-order in the QD-phonon coupling.
Unlike the polaronic approaches above, electron-phonon interactions were included only to first order, which is generally not valid
in these cavity-QED systems---even at low temperature~\cite{roy_hughes} (see Sec.~\ref{multiphoton_section}).
More problematic is the fact that the effects of the coherent drive on the phonon reservoir and the
 associated EID effects are missing; in contrast, we find these to be the dominant source of broadening from
 electron-phonon scattering.
   The need for  EID processes in coherently driven semiconductor cavity-QED system has already been shown for the  Mollow triplet, both
   experimentally~\cite{stuttgart_prl}  and theoretically~\cite{roy_hughes}.
%

In addition to the
phonon-induced  Lindblad decay rates above,
one also has phonon-mediated frequency shifts beyond the polaron shift.
The effective Hamiltonian, describing the coherent part of the system evolution, $H_{\rm sys}^{\rm eff}$, becomes
\begin{align}
\label{sec3eqfinal5a}
H^{\rm eff}_{sys}& = \hbar\Delta_{xL}\hat{\sigma}^{+}\hat{\sigma}^{-}
+\hbar\Delta_{cL} \hat{a}^{\dagger}\hat{a}+\langle B\rangle \hat{X}_{g} \nonumber \\
& + \hbar\Delta^{\sigma^{+}a}_{ph}\hat{a}^{\dagger}\hat{\sigma}^{-}\hat{\sigma}^{+}\hat{a} + \hbar\Delta^{a^{\dagger}\sigma^{-}}_{ph}\hat{\sigma}^{+}\hat{a}\hat{a}^{\dagger}\hat{\sigma}^{-} \nonumber \\
& + \hbar\Delta^{\sigma^{-}}_{ph}\hat{\sigma}^{-}\hat{\sigma}^{+} + \hbar\Delta^{\sigma^{+}}_{ph}\hat{\sigma}^{+}\hat{\sigma}^{-},
\end{align}
with
\begin{align}
\label{sec3eqfinal5b}
\Delta_{ph}^{\sigma^{-}/\sigma^{+}}&= \braket{B}^2\!\eta_{x}^{2}\,{\rm Im}\left [\int_{0}^{\infty}d\tau\,
e^{\pm i\Delta_{xL}\tau} \left (e^{\phi(\tau)}-1\right )\right], \\
\Delta_{ph}^{\sigma^{+}a/a^{\dagger}\sigma^{-}} & = \braket{B}^2\!g^{2}\,{\rm Im}\left [\int_{0}^{\infty}d\tau\,
e^{\pm i\Delta_{cx} \tau}\left (e^{\phi(\tau)}-1\right )\right], \ \ \
\end{align}
where $\Delta_{ph}^{\sigma^{+}a}$, $\Delta^{a^{\dagger}\sigma^{-}}_{ph}$, $\Delta_{ph}^{\sigma^{+}}$ and $\Delta^{\sigma^{-}}_{ph}$
 are the 
Stark shifts (which scale proportionally with $\braket{B} g^2$ or $\braket{B} \eta_x^2$).

\section{Numerical Results}
\label{results}

\subsection{Role of Phonon Scattering on Intensity Power Broadening:\
Effective Phonon ME Versus the Full Time-Convolutionless  ME}

\begin{figure}[t!]
\vspace{-0.1cm}
\centering\includegraphics[width=1\columnwidth]{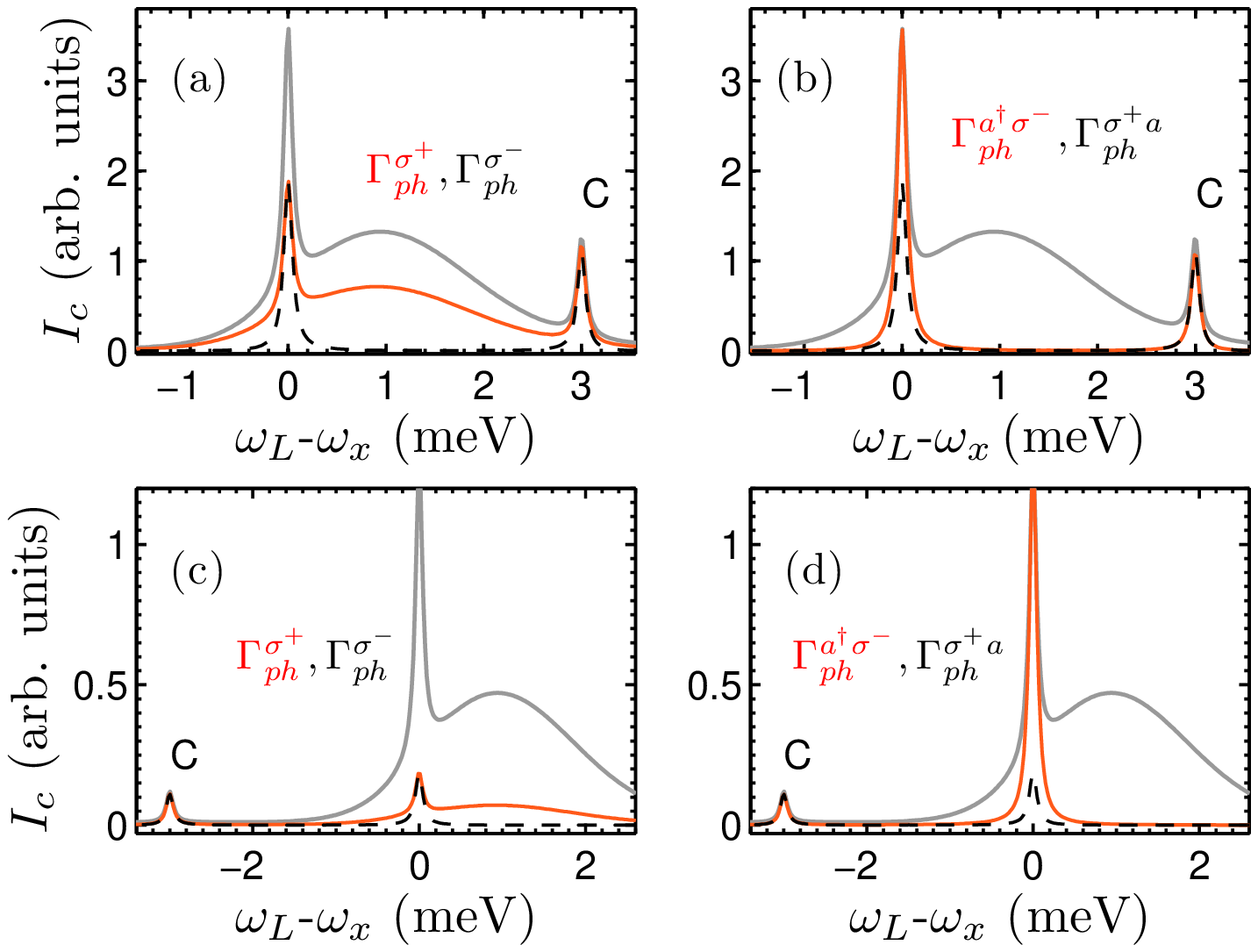}
\vspace{-0.3cm}
\caption{(Color online) (a) Influence of the effective phonon scattering rates, $\Gamma_{ph}^{\sigma^{+}}$ (red, lower solid line)
and $\Gamma_{ph}^{\sigma^{-}}$ (black, dashed line) [defined in Eq.~(\ref{sec3eqfinal5})],
on the intensity PL, $I_c$, for a QD driven system at $T=4$~K with $\Delta_{cx}=3$ meV. Also shown is the full polaron ME solution (grey, upper solid line) [Eq.~(\ref{sec3eq3})]. We plot the cavity intensity ($I_c$) for exciton excitation as a function of QD-laser detuning and show the contribution of one effective Lindblad 
 rate per calculation;
  the collective influence  of these processes is shown later in Fig.~\ref{fig:5n} and compared with the
  full solution. (b) As in (a), but for
 effective phonon scattering rates, $\Gamma_{ph}^{a^{\dagger}\sigma^{-}}$ (red, lower solid line) and
 $\Gamma_{ph}^{\sigma^{+}a}$ (black, dashed line) [defined in Eq.~(\ref{sec3eqfinal5R})].  The system and material parameters are: $\gamma=2~\mu$eV, $\kappa=50~\mu$eV, $g=20~\mu$eV, $\eta_x=40~\mu$eV, $\gamma^{\prime}(2\,{\rm K})=2~\mu$eV,
  and we compute $\braket{B}(4\,{\rm K})=0.91$.
   (c-d) As in (a-b) but with
   $\Delta_{cx}=-3$ meV.  }
 \label{fig:3n}

\vspace{0.3cm}
\includegraphics[width=1\columnwidth]{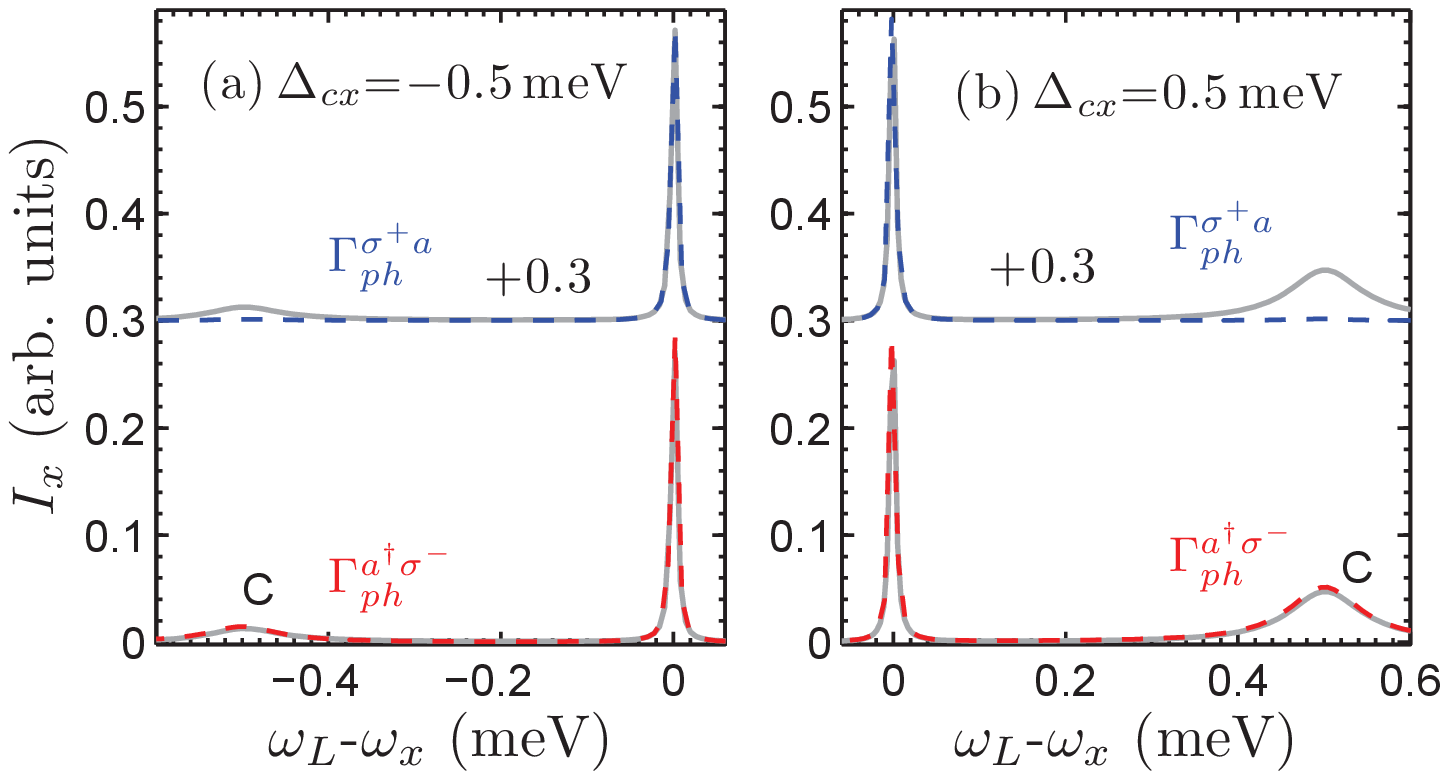}
\vspace{-0.3cm}
\caption{(Color online) (a) Influence of the role of the effective phonon scattering rates,
 $\Gamma_{ph}^{a^{\dagger}\sigma^{-}}$  (red, lower, dashed line)
and $\Gamma_{ph}^{\sigma^{+}a}$ (blue, upper, dashed line),  on the intensity PL, $ I_x$, for a cavity driven system at $T=4$~K for $\Delta_{cx}=-0.5$ meV. Also shown is the full
polaron ME solution
(solid grey lines) [Eq.~(\ref{sec3eq3})].
For clarity, the $\Gamma_{ph}^{\sigma^{+}a}$ curves are shifted vertically by 0.3, along with the full polaron ME solution.
We plot the contribution of one effective Lindblad rate per calculation.
(b) As in (a) but for $\Delta_{cx}=0.5$ meV.
The  parameters are
  the same as  in Fig.~\ref{fig:3n}, but with
   $\eta_c=40~\mu$eV (cavity pumping).
}
 \label{fig:4n}
 \vspace{-0.4cm}
\end{figure}

We first investigate the role of the four phonon Lindblad terms in a typical power-broadened intensity PL computed with our
EPME [Eq.~(\ref{effectiveME})] and compare with the full solution (Eq.~(\ref{sec3eq3}): time-convolutionless  ME).
The main parameters are listed in Fig.~\ref{fig:3n}, and we have 
adopted  system parameters and coupling constants similar to those in recent semiconductor
 experiments~\cite{jelena1,ates3}.

In Fig.~\ref{fig:3n} we study the role of $\Gamma_{ph}^{\sigma^{-}}$, $\Gamma_{ph}^{\sigma^{+}}$, $\Gamma_{ph}^{a^{\dagger}\sigma^{-}}$ and $\Gamma_{ph}^{\sigma^{+}a}$, as defined in Eqs.~(\ref{sec3eqfinal5}-\ref{sec3eqfinal5R}), on the power broadening lineshape
 ($I_c \propto \bar n_c$); here we use  $\eta_x=40\,\mu$eV for a QD driven system at a bath temperature of $T=4$~K,
 and study two different cavity-exciton detunings, (a-b) $\Delta_{cx}=3$ meV and (c-d) $\Delta_{cx}=-3$ meV.
 The corresponding peak $\bar n_c$ that results from this interaction is  around $4\times 10^5$. To better highlight the various scattering mechanisms, we include only one of the Lindblad terms in each calculation, 
 as labeled in the plots.
By looking at Fig.~\ref{fig:3n}(a,c), it is clear that
 the process $L(\hat{\sigma}^{+})$ is primarily responsible for {\em incoherently} exciting the phonon sidebands [cf.~\ref{fig:3n}(a)]
 and EID; while process $L(\hat{\sigma}^{-})$ introduces further pump-dependent EID, as will be highlighted
 in detail later
 (note that this process has the same Lindblad operator terms as $\gamma$). The broad background centered at
$\omega_L-\omega_x = 1$ meV is present only for the case of $\Gamma^{\sigma^+}$. The exciton-cavity scattering processes,
$L(\hat{a}^{\dagger}\hat{\sigma}^{-})$  and $L(\hat{\sigma}^{+}\hat a)$, account for cavity excitation
 and cavity destruction, respectively, by phonon-assisted processes and these
 affect the relative magnitudes of the cavity measured intensity PL at different temperatures and drives.
Figures~\ref{fig:3n}(b,d)  demonstrate
that $L(\hat{a}^{\dagger}\hat{\sigma}^{-})$  is the main cavity-exciton coupling ({\em feeding}) term;
this mechanism results in enhanced cavity photon numbers at the exciton transition,
especially when the cavity is red shifted
from the exciton---since phonon emission is favorable at lower
temperatures. In contrast, the $L(\hat{\sigma}^{+}\hat a)$ process gives no 
noticeable exciton-cavity coupling
because of the fast cavity decay rate~\cite{HohenesterPRB:2010}. 
%

\begin{figure}[t!]
\centering\includegraphics[width=1\columnwidth]{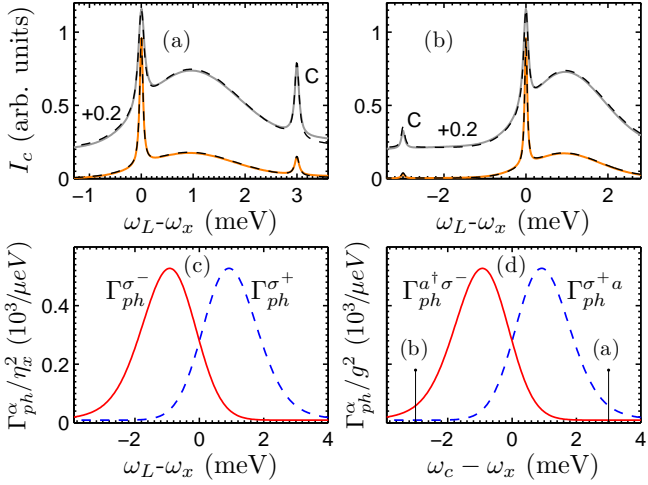}
\caption{(Color online) Phonon bath at $T=4$~K.\ (a-b) Normalized cavity mode intensity ($I_c$) for a dot-driven system as a function of QD-laser detuning  for two different dot-cavity detunings, (a) $\Delta_{cx} =3$ meV and (b) $\Delta_{cx}=-3$~meV, and for two different values of the cw laser Rabi frequency (orange, lower solid line, corresponds to $\eta_{x}=20~\mu$eV; and grey,
upper solid curve, corresponds to $\eta_{x}=40~\mu$eV). Also shown (black, dashed lines) are the intensity PL obtained using the effective Lindblad form of the full time-convolutionless ME. Note that we have vertically shifted the intensity PL for $\eta_{x}=40~\mu$eV by $0.2$ for clarity.
 (c) Plot of the phonon rates $\Gamma_{ph}^{\sigma^{+}/\sigma^{-}}$ (see text) for $\Delta_{cx}=-3$~meV as a function of QD-laser detuning.
%
  Note $\Gamma_{ph}^{\sigma^{-}}(-\Delta_{cx})=\Gamma_{ph}^{\sigma^{+}}(\Delta_{cx})$.
  (d) Plot of the phonon rates, $\Gamma_{ph}^{\sigma^{+}a/a^{\dagger}\sigma^{-}}$, as a function of QD-cavity detuning. 
}
  \label{fig:5n}
\vspace{0.0cm}
\end{figure}

In Fig.~\ref{fig:4n} we carry out a similar exercise for
a  cavity-excited system ($\eta_c=40~\mu$eV), calculating $I_x$ ($\propto \bar n_x$),
where we study the influence
of $L(\hat{a}^{\dagger}\hat{\sigma}^{-})$ and $L(\hat{\sigma}^{+} \hat a)$
on power broadening;
here we find excellent agreement with only the
$\Gamma_{ph}^{ a^\dagger  \sigma^{-}}$ scattering term (lower red, dashed curve) compared to the full
ME solution (lower grey, solid line), which results in a significant exciton-cavity feeding process via phonon emission [cavity is red detuned
in (a), cf.~Fig.~\ref{fig:2n}(d)]. Again we find that $L(\hat{\sigma}^{+}\hat a)$ gives no noticeable
cavity feeding. For this cavity driven system, 
we have chosen $\Delta_{cx} =\pm 0.5$ meV instead of 
 $\Delta_{cx} =\pm 3$ meV (which we chose earlier for the exciton driven system);
 this is because a strong exciton-driven system invariably kicks up the phonon sidebands even at low temperatures 
 that can swamp the emission at the cavity mode; so we use a larger cavity-exciton detuning for the exciton driven case. The cavity driven system is thus much {\em cleaner} to analyze
 for smaller cavity-exciton detunings,
 and the exciton-measured intensity PL is also substantially reduced 
 for larger detunings.
Note that the corresponding peak $\bar n_x$ for this excitation regime is  around $0.3.$

\begin{figure}[t!]
\centering\includegraphics[width=1\columnwidth]{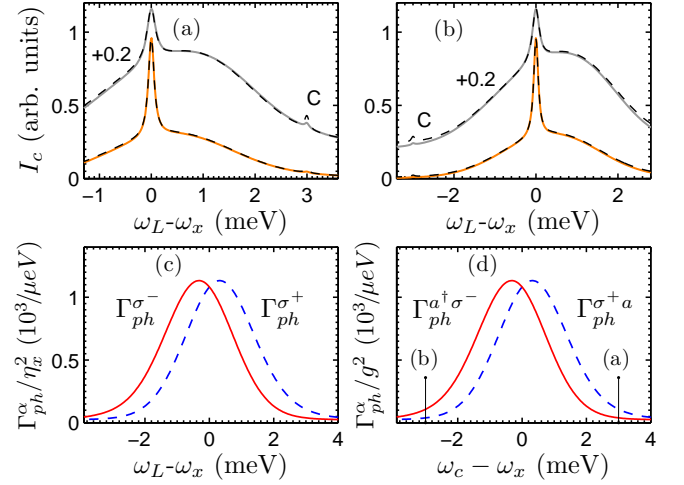}
\caption{(Color online)
As in Fig.~\ref{fig:5n} but with the phonon bath at $T=20$~K.
The system parameters are identical to those given in Fig.~\ref{fig:3n},
except that
$\gamma^{\prime}(20\,{\rm K})=20~\mu$eV
  and we now compute $\braket{B}(20\,{\rm K})=0.73$ (cf.~$\braket{B}(4\,{\rm K})=0.91$).}
\label{fig:6n}
\end{figure}

In Figs.~\ref{fig:5n}(a,b)  we plot the normalized cavity mode intensity, $I_c \propto \bar n_c$, for a dot-driven system as a function of QD-laser detuning, again for the two different cavity-exciton detunings, (a) $\Delta_{cx}=3$ meV and (b) $\Delta_{cx}=-3$ meV,  
at $T=4$~K.
We also show the intensity PL for   $\eta_x=20~\mu$eV and $\eta_x=40~\mu$eV.
To study the effects of increasing temperatures,
in Fig.~\ref{fig:6n} we  plot the
$I_c$ at $T=20$~K. In these graphs,
we 
show the 
{\em total} 
power broadened intensity PL obtained using the EPME [dashed lines in (a-b)]
and compare with the full polaron ME solution [solid lines in (a-b)]; given
the approximations made in the derivation of the EPME, the agreement is remarkably good. The frequency-shift terms, $\Delta_{ph}^\alpha$, are found to be very small here and can be neglected for the cases shown. Note that the phenomenological pure-dephasing rates are
 temperature dependent, where we choose $\gamma^{\prime}(4\,{\rm K})=2~\mu$eV and $\gamma^{\prime}(20\,{\rm K})=20~\mu$eV (e.g., see Refs.~\onlinecite{BorriPRL:2001,ota}).
 The thermal expectations of the phonon displacement operators are calculated to be $\braket{B}(4\,{\rm K})=0.91$ and $\braket{B}(20\,{\rm K})=0.73$. These results suggest that the dynamics can be well described by our EPME,  by essentially 
only including three separate phonon scattering effects---since, from the findings above,
$L(\hat{\sigma}^{+}\hat a)$ can be safely neglected.

In Fig.~\ref{fig:5n}(c)   we plot the corresponding phonon scattering  rates, $\Gamma_{ph}^{\sigma^{-}}$ and $\Gamma_{ph}^{\sigma^{+}}$, as a function of QD-laser detuning for
$\eta_{x}=40~\mu$eV, at $T=4$~K; in
 Fig.~\ref{fig:6n}(c) we plot these scattering rates
  for  $T=20$~K.
  Since the rates depend on QD-laser detuning and the pump strength, they are
  obviously important for understanding power broadening in a cavity-QED system.
  In Figs.~\ref{fig:5n}-6(d)  we plot the rates, $\Gamma^{\sigma^{+}a}$ and $\Gamma^{a^{\dagger}\sigma^{-}}$, 
  as a function of QD-cavity detuning for two different temperatures;  
   the regions marked by the vertical lines indicate the chosen detunings in
   Figs.~(a) and (b)---so note that these particular rates are fixed as a function of QD-laser detuning.
   As discussed earlier,
   the scattering term, $\Gamma^{\sigma^{+}a}$, describes a process which involves de-exciting the QD and exciting the cavity mode, aided through phonon emission or absorption.
  
The collective influence from the various
  phonon scattering terms, discussed above,
  results in  broadening (EID) of the QD exciton resonance, {\em incoherent} excitation
 of the phonon bath, and significant exciton-cavity coupling (or feeding); the first
  two of these are pump-dependent, through $\sim \braket{B}^2\eta_x^2$, while
  the latter process scales  with $\braket{B}^2g^2$.
 The trends
of the exciton-cavity feeding rate, $\Gamma^{a^\dagger\sigma^-}_{ph}$,
are consistent with 
the results of Refs.~\onlinecite{JiaoJPC2008,HohenesterPRB:2009},
 where one observes a peak 
 scattering rate at around
 $\Delta_{cx} \sim 1-2~$meV (depending upon the temperature).
  Note that at $T=4$~K and $\Delta_{cx}= -3$~meV, $\Gamma^{\sigma^{+}a}$ dominates, whereas $\Gamma^{a^\dagger\sigma^{-}}$ is much larger for $\Delta_{cx}= 3$~meV. However, at $T=20$~K, the two rates are much closer to each other for the different detunings as the dependence of the rates on QD-cavity detuning becomes more symmetric with increasing temperature~\cite{HohenesterPRB:2009}.


\begin{figure}[t!]
\centering\includegraphics[width=1\columnwidth]{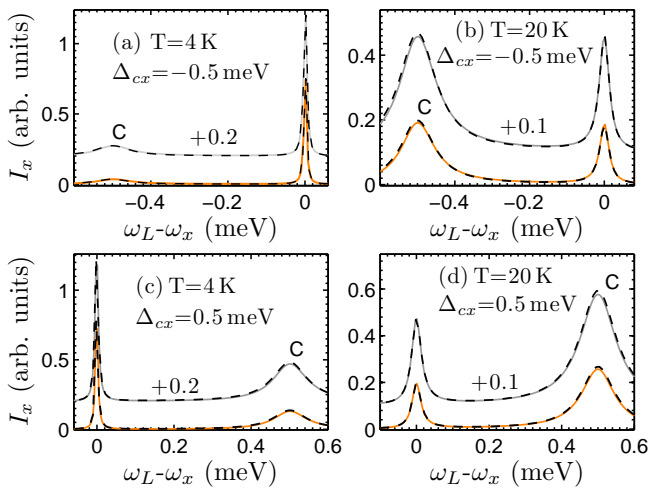}
\caption{(Color online)
(a) and (b) Relative QD intensity for a cavity-driven system as a function of QD-laser detuning for 
$\Delta_{cx} =-0.5$ meV and phonon bath temperature (a) $T=4~$K  and (b) $T=~20$~K and for two different values of the
  cw laser Rabi frequency (orange, lower, solid line corresponds to $\eta_{c}=40~\mu$eV, and grey, lower, solid line corresponds to $\eta_{c}=60~\mu$eV). Also shown (black dashed curves) are the intensity PL obtained using the effective Lindblad form of the full time-convolutionless ME. The parameters were as follows: $\gamma_{x}=2~\mu$eV, $\kappa=50~\mu$eV, $\gamma^{\prime}=2~\mu$eV at $T=4$~K, and $\gamma^{\prime}=20~\mu$eV at $T=20$~K. (c)  Same as in (a) and (b) but with $\Delta_{cx}=0.5$~meV.}
\label{fig:7n}
\end{figure}

  It is interesting to note that the broadening of the QD
  exciton  resonance as a function of QD-laser detuning closely mirrors the PL lineshape associated with the 
  linear exciton spectrum obtained
  using the IBM. A similar observation was  demonstrated by Ahn {\em et. al.}~\cite{knorr2}, where the effects of electron-phonon coupling in QDs (with no cavity) on nonstationary resonance fluorescence spectra were studied; the resonance fluorescence dynamics of the QD electronic transition was shown to have a strong dependence on the duration of the laser field, and  by increasing duration of the laser pulse, the background phonon continuum was strongly excited.

Similar to the QD driven system,
 in Fig.~\ref{fig:7n} we  also investigate the {exciton} PL characteristics, $I_x \propto \bar n_x$,
 but with {\em cavity} excitation
 (investigated in more detail in Subsect.~\ref{C:Pumped}).
As anticipated from Fig.~\ref{fig:4n},
we again obtain
a very good fit between our EPME results and the full polaron time-convolutionless ME solution. In general, the effective phonon Lindblad solution is
expected to
closely mimic the full solution here for large QD-cavity detunings ($|\Delta_{cx}| \gg g,\eta_c$), though we find that it can
work 
well over a wide range of excitation conditions.


\begin{table}[b!]
\caption{\label{tab:example1}Dependence on the $x/c$ FWHM of the intensity PL on $\gamma$, for 
a QD-driven cavity measured PL ($x$ FWHM, with $\Delta_{cx}=3~$meV), and for a cavity-driven
  QD measured PL
  ($c$ FWHM, with
$\Delta_{cx}=0.5~$meV). The QD/cavity parameters are $\kappa=50~\mu$eV, $g=20~\mu$eV, $\eta_{(x/c)}=30~\mu$eV,
and $\gamma^{\prime}=2~\mu$eV (4~K). No coupling to phonons is considered 
and all numbers are in units of $\mu$eV.}
\begin{ruledtabular}
\begin{tabular}{lll}
 \vspace{0.18cm}
 & $x$ FWHM (QD driven) & $c$ FWHM (cavity driven)\\
$\gamma$ = 2  & 120 & 107\\
$\gamma$ = 4  & 104 & 107\\
$\gamma$ = 6  & 98 & 107\\
\end{tabular}
\end{ruledtabular}
\end{table}
\begin{table}[b!]
\vspace{-0.2cm}
\caption{\label{tab:example2}Dependence on the $x/c$ FWHM of the power broadened lineshape on $\kappa$, for 
a QD-driven cavity measured system  
 ($x$ FWHM, with $\Delta_{cx}=3~$meV), and a for cavity-driven QD measured PL
($c$ FWHM, with $\Delta_{cx}=0.5~$meV). The QD/cavity parameters 
 are the same as in Table~\ref{tab:example1},
and all units are in $\mu$eV.}
\begin{ruledtabular}
\begin{tabular}{lll}
 \vspace{0.18cm}
 & $x$ FWHM (QD driven) & $c$ FWHM (cavity driven)\\
$\kappa$ = 10  & 120 & 41\\
$\kappa$ = 30  & 120 & 60\\
$\kappa$ = 50  & 120 & 107\\
\end{tabular}
\end{ruledtabular}
\end{table}
\begin{table}[b!]
\vspace{-0.2cm}
\caption{\label{tab:example3}Dependence on the $x/c$ FWHM of the power broadened lineshape on $\gamma^{\prime}$, for 
 a QD-driven cavity measured PL system ($x$ FWHM, with $\Delta_{cx}=3~$meV),  and for a 
 cavity-driven  QD measured PL ($c$ FWHM,
with $\Delta_{cx}=0.5~$meV). The QD/cavity parameters
are the same as in Table~\ref{tab:example1}, and
all units are in $\mu$eV.}
\begin{ruledtabular}
\begin{tabular}{lll}
 \vspace{0.18cm}
 & $x$ FWHM (QD driven) & $c$ FWHM (cavity driven) \\
$\gamma^{\prime}$ = 2  & 120 & 107\\
$\gamma^{\prime}$ = 4  & 146 & 107\\
$\gamma^{\prime}$ = 6 & 169 & 107\\
\end{tabular}
\end{ruledtabular}
\end{table}

\begin{table}[t!]
\vspace{0.1cm}
\caption{\label{tab:example4} Dependence on the $x$ FWHM of the power broadened intensity PL of a
QD-driven cavity measured  system ($\Delta_{cx}=3~$meV), on the two phonon Lindblad processes, $\Gamma_{ph}^{\sigma^{+}}$ and $\Gamma_{ph}^{\sigma^{-}}$, at $T=4$~K,
for various values of $\eta_x$. 
The general system parameters
 are the same as in Table~\ref{tab:example1}.
Also considered is the case of no incoherent phonon coupling  with $\braket{B}=1$ (i.e., also no coherent phonon
coupling),
and $\braket{B}=0.91$ at $T=4$~K (with coherent phonon coupling).
 The inclusion of  $\braket{B}$ at 4~K here is to make a better comparison
 with the same effective $\eta_x$ ($\rightarrow \braket{B}\eta_x$) and effective $g$ ($\rightarrow\braket{B}g$); 
 in reality, the no-phonon case
 will be different  because of the lack of coherent phonon renormalizations (later we will show
 this difference more explicitly, e.g., in Figs.~\ref{fig:10n}-\ref{fig:13n}).
 All numbers are in units of $\mu$eV, and note that the broadening here
 is independent of 
  $\Gamma_{ph}^{\sigma^{+}a}$ and $\Gamma_{ph}^{a^{\dagger}\sigma^{-}}$.
  The last column gives the full polaron ME result [Eq.~(\ref{sec3eq3})]}
\begin{ruledtabular}
\begin{tabular}{lllllll}
& No $\Gamma_{ph}^\alpha$ & No $\Gamma_{ph}^\alpha$ &
$\Gamma_{ph}^{\sigma^{+}}$ & $\Gamma_{ph}^{\sigma^{-}}$ & EPME & Full \\
 \vspace{0.18cm}
 & $\braket{B}=1$ & $\braket{B}$(4~K) &  &  & &\\
$\eta_{x}$ = 20 & 80 & 73 & 71 & 72 & 70 & 72\\
$\eta_{x}$ = 40 & 159 & 146 & 138 & 139 & 134 & 151\\
$\eta_{x}$ = 60 & 240 & 219 & 251 & 200 & 223 & 390\\
\end{tabular}
\end{ruledtabular}
\vspace{0.1cm}
\caption{\label{tab:example5}Dependence on the $c$ FWHM of the power broadened intensity PL of a 
 cavity-driven QD measured system, on the 
two phonon Lindblad processes, $\Gamma_{ph}^{\sigma^{+}a}$ and $\Gamma_{ph}^{a^{\dagger}\sigma^{-}}$, at $T=4$~K, for various values of $g$.
Also considered is the case of no phonon coupling,
with two $\braket{B}$ values as above. 
The general system parameters
are the same as in Table~\ref{tab:example1}, and
all numbers are in units of $\mu$eV.}
\begin{ruledtabular}
\begin{tabular}{lllll}
& No $\Gamma_{ph}^\alpha$ &No $\Gamma_{ph}^\alpha$ &$\Gamma_{ph}^{\sigma^{+}a}$ & $\Gamma_{ph}^{a^{\dagger}\sigma^{-}}$\\
\vspace{0.18cm}
& $\braket{B}=1$ & $\braket{B}$(4~K) &  & \\
g = 20 & 101 & 101  & 101 & 102\\
g = 40 & 101 & 101  & 101 & 109\\
g = 60 & 101  & 101  & 101 & 115\\
\end{tabular}
\end{ruledtabular}
\vspace{0.1cm}
\caption{\label{tab:example6}Dependence on the  $c$ FWHM of the power broadened intensity PL of a  cavity-driven 
QD measured
system ($\Delta_{cx}=0.5~$meV), on the two Lindblad processes, $\Gamma_{ph}^{\sigma^{+}a}$ and $\Gamma_{ph}^{a^{\dagger}\sigma^{-}}$, at $T=4$~K, for various values of $\eta_c$.
Also considered is the case of no phonon coupling  with two
 $\braket{B}$  values. 
 The general system parameters
are the same as in Table~\ref{tab:example1}, and
 all numbers are in units of $\mu$eV. }
\begin{ruledtabular}
\begin{tabular}{lllllll}
 & No $\Gamma_{ph}^\alpha$ & No $\Gamma_{ph}^\alpha$ & $\Gamma_{ph}^{a^{\dagger}\sigma^{-}}$
 & $\Gamma_{ph}^{\sigma^{+}a}$   & EPME & Full \\
 \vspace{0.18cm}
 & $\braket{B}=1$ & $\braket{B}$(4~K) &  &  & &\\
$\eta_{c}$ = 20 &  100 & 100 & 100 & 100 & 100 & 100\\
$\eta_{c}$ = 40 & 101 & 101 & 101 & 103 & 103 & 103\\
$\eta_{c}$ = 60 & 102 & 102 & 102 & 107 & 108 & 108\\
\end{tabular}
\end{ruledtabular}
\end{table}


A preliminary  analysis of the role of the
various scattering processes
on the intensity PL is presented  in Tables~\ref{tab:example1}-\ref{tab:example6}.
Here we  focus on the qualitative influence of the various scattering rates
on the exciton ($x$) and cavity ($c$) FHWM broadenings, but later we will investigate the full 
power-broadened PL curves in detail.
In Table~\ref{tab:example1}
we study the dependence of the $x/c$ FWHM
of the intensity PL on $\gamma$ (the radiative decay rate). Increasing $\gamma$ results in reduced broadening of a
cavity-measured QD driven PL
lineshape ($x$ FWHM). Increasing $\gamma$ broadens the QD resonance resulting in reduced coupling between the cavity mode and the QD and subsequent narrowing of the cavity-measured QD driven PL lineshape. However, $\gamma$ has little effect on the power broadening of a
QD-measured cavity driven intensity PL
($c$ FWHM). In Table~\ref{tab:example2} we present numerical results for the $x/c$ FWHM of the
 intensity PL for different values of $\kappa$. Increasing $\kappa$ here has no
observable effect on
QD-driven  cavity measured 
intensity PL,  but increasing $\kappa$ results in increased broadening of a
 cavity-driven QD measured 
intensity PL.  In Table~\ref{tab:example3} we study the dependence on $\gamma^{\prime}$, i.e., pure dephasing of the ZPL.
Increasing $\gamma^{\prime}$ increases broadening of  QD-driven  cavity measured 
intensity PL, but
has negligible effect on the cavity driven  QD-measured system.

In Table~\ref{tab:example4}
we study the role of the two phonon Lindblad processes,
$\Gamma_{ph}^{\sigma^{+}}$ and $\Gamma_{ph}^{\sigma^{-}}$, on a  QD- driven cavity-measured system at $T=4$~K, for
various values of $\eta_x$; the other two phonon
processes, i.e., $\Gamma_{ph}^{\sigma^{+}a}$ and
$\Gamma_{ph}^{a^{\dagger}\sigma^{-}}$, do not influence the power broadening lineshape here. As can be seen,
increasing the influence of these effective Lindblad processes by increasing $\eta_x$ reduces the
linewidth of the cavity measured intensity PL.
We also see that the $x$ FWHM broadening for increasing drives  may be drastically
underestimated by the effective phonon ME, even though the qualitative
trends of the full intensity PL are similar; this is because
we have neglected the influence of the drive on the 
phase integrations that enter the full incoherent phonon scattering
[see Eqs.~(\ref{sec3eq2b}) and (\ref{sec3eq3})].

In Table~\ref{tab:example5}, we 
list the $c$ FWHM of the power-broadening lineshape of a  cavity-driven QD measured system on the two Lindblad processes,
$\Gamma_{ph}^{\sigma^{+}a}$ and $\Gamma_{ph}^{a^{\dagger}\sigma^{-}}$, at $T=4$~K, for various values
of $g$. The Lindblad process caused by $\Gamma_{ph}^{\sigma^{+}a}$ has no effect on the $c$ FWHM. However, increasing
$\Gamma_{ph}^{a^{\dagger}\sigma^{-}}$ (the cavity feeding process) results in some increased broadening. Finally, in
Table~\ref{tab:example6} we
show the influence of these phonon processes 
($\Gamma_{ph}^{a^{\dagger}\sigma^{-}/\sigma^+a}$)
with increasing
$\eta_c$, which also show a slight increase of the $c$ FWHM  with increasing drives.
Now we see that the $c$ FWHM broadening for increasing drives
is extremely well reproduced with the effective phonon ME; this is because the cavity
pump no longer enters $H'_{sys}$ [see Eqs.~(\ref{sec3eq2b1}) and (\ref{sec3eq3})]. 
In both Tables~\ref{tab:example5}-\ref{tab:example6},
we see no role from the slight coherent reduction
of $\braket{B}\!g$ at 4~K.
In general we see that driving via the exciton or the cavity
can yield drastically different results.


To summarize this subsection, we have identified
three main Lindblad phonon processes that contribute to
the intensity PL of a cavity-QED system. 
We have also found earlier that 
the effective scattering rates
associated with these processes can be calculated, under certain
detuning conditions, from  
simple
analytical solutions  [Eqs.~\ref{sec3eqfinal5}-\ref{sec3eqfinal5R}];
these 
{\em effective} Lindblad solutions are compared to the full polaron ME solution
and found, in certain cases, to yield very good agreement. In this way we can also argue
the underlying physics of the identified phonon scattering processes.
However, there can be noticeable differences, especially for the measured
$x$ FWHM broadening of an exciton driven system.

In what follows below,  we will use the full polaron ME, and first verify the
general need for for multiphonon and mutiphoton effects.

\subsection{Influence of Multiphonons and  Multiphotons on the Intensity PL}
\label{multiphoton_section}

 Our time-convolutionless ME  above [Eq.~(\ref{sec3eq3})] utilizes the polaron frame which allows
 for a nonperturbative treatment of phonons. This enables one to use the full IBM machinery to compute the phonon
 correlation functions. It is also useful to look at the one phonon limit of the time convolutionless ME
 by expanding the phonon Green functions to lowest order in the phonon coupling. In this limit we can expand the phonon
 correlation function $\phi(t)$ to lowest order in the dot-phonon coupling constant
 as follows: $G_{g}(t)\simeq 0$ and $G_{u}(t)\simeq\phi(t)$ where we have used $\braket{B}\simeq 1$.
 This then connects to a weak-phonon coupling (i.e., perturbative) approach.

\begin{figure}[t!]
\centering\includegraphics[width=0.99\columnwidth]{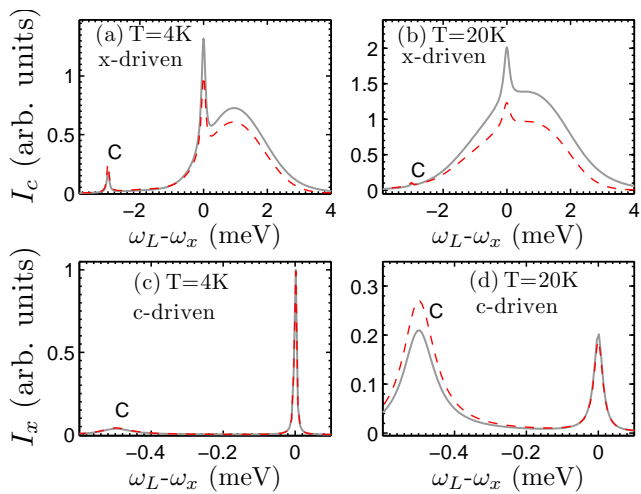}
\caption{(Color online) (a-b) Cavity mode intensity PL for a QD driven system ($\eta_{x}=30~\mu$eV) as a function of QD-laser detuning for a one-phonon theory (red dashed line) and the full polaron model (grey solid line) with $\Delta_{cx} =-3$ meV, for phonon bath tempereatures (a) $T=4$~K and (b) $T=20$~K. (c-d) Relative QD (exciton) intensity for a cavity-driven system ($\eta_{c}=30~\mu$eV) as a function of QD-laser detuning for a one-phonon theory (red dashed line) and the full polaron model (grey solid line) with $\Delta_{cx} =-0.5$ meV, for phonon bath temperatures (c) $T=4$~K and (d) $T=20$~K.}
\label{fig:8n}
\end{figure}

 In Figs.~\ref{fig:8n}(a) and (b)  we plot the cavity mode intensity ($I_c$) for an exciton driven system as a function of
 exciton-laser detuning for a one phonon and the full polaron (multi-phonon) solution. We recognize that
the one phonon scattering process tends to deviate from the full polaron intensity PL, especially
noticeable at higher temperatures and for larger driving field strengths.
 Thus even for low temperatures, a weak-phonon coupling theory can break down.
 We further find that the one phonon approximation {overestimates} the power broadening intensity PL for a
 negatively detuned cavity.  Similar conclusions 
 about the need for multiphonon effects were also found
 in the context of the resonance fluorescence spectra of a QD driven cavity-QED system~\cite{roy_hughes} and
 time-dependent excitonic Rabi rotation dynamics~~\cite{nazir2}.
However, for a cavity driven system [Fig.~\ref{fig:8n}(c) and (d)], at  low temperatures (e.g.,
4~K), the 
weak phonon theory can be accurate.
Of course, other QD material parameters
can yield different trends in the role
and assessment of multiphonon coupling.
Since the phonon correlation functions that we use are well known for an IBM model, the full
 phonon calculation of the intensity PL presents the same level of computational complexity as the
 one phonon calculations---which is a major advantage of the ME formalism above. 


\begin{figure}[t!]
\centering\includegraphics[width=1\columnwidth]{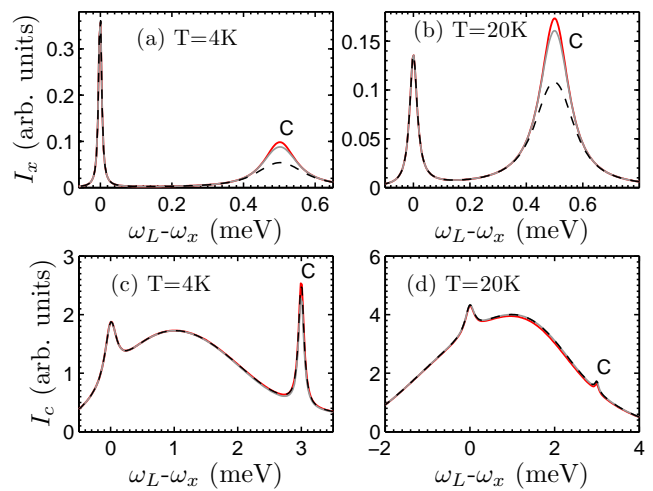}
\caption{(Color online) (a-b) Exciton intensity PL for a cavity driven system ($\eta_c=30~\mu$eV) as a function of QD-laser detuning (with $\Delta_{cx} =0.5$ meV), for different numbers of truncated photon states at (a) $T=4$~K and (b) $T=20$~K: two-photon (or more, since this value has converged): black
dashed line,
four-photon: grey solid, lighter line, six-photon (or more): red solid, darker line.  (c-d) Mean cavity photon number for a dot driven system as a function of QD-laser detuning (with $\Delta_{cx} =3$~meV), for different number of truncated photon states with (c) $T=4$~K and (d) $T=20$~K: one-photon: black dashed line, two-photon: grey, lighter solid line,
and three-photon: red solid, darker line.}
\label{plot_exp_4_fig}
\label{fig:9n}
\end{figure}
We also study the influence of quantized {\em multiphoton} processes in the intensity PL of both an
 exciton-driven and a cavity-driven system. In Figs.~\ref{fig:9n}(a)and \ref{fig:9n}(b), using
  cavity excitation ($\eta_{c}=30~\mu$eV), we plot  $I_x$
  as a function
  of QD-laser detuning for various truncations of the photon Hilbert state.
  We find that it is necessary to include up to  six-photon processes (e.g., {\em a 13 state model}) to correctly
  describe the effect of cavity photons; note that including more than six photon processes yields an identical result to the six photon calculations, so these values have converged on the numerically exact answer. 
  While the need for sox photon processes may seem surprising for the relatively small values of $g$ (i.e., $20~\mu$eV), the exciton-phonon
  coupling with phonon scattering can be more sensitive to quantum cavity-QED
  effects.
  In the rest of the paper, we thus compute the intensity PL
  of the cavity driven system  in a six-photon truncated basis.

  In Fig.~\ref{fig:9n}(c) and
\ref{fig:9n}(d) we plot the  cavity intensity PL ($I_c$) for a QD driven system ($\eta_{x}=30~\mu$eV), as a function of
exciton-laser detuning for various truncations of the photon Hilbert state. Unlike a cavity driven system, we now find
that a two-photon truncation of the photon Hilbert space is enough to obtain a correct (i.e., converged in the photon basis) description of the cavity PL. The fundamental difference between a QD-driven and a cavity-driven system is due to the fact that the
cavity is represented as a quantized harmonic oscillator whereas a QD is
a two-level system. Hence, for our system parameters, the cavity is more easily excited into higher lying Jaynes-Cummings
ladder states, even though the system exhibits relative strong dissipation (i.e., $\kappa = 2.5 g$).
Neglecting quantum multiphoton processes can therefore introduce
spurious effects in the PL lineshape.
 A two-photon truncation was also found to be
sufficient (and necessary) for the study of the fluorescence spectrum of a QD driven cavity-QED system~\cite{roy_hughes}.

In summary,
 the results above highlight the need for both multiphonon and multiphoton effects
 for understanding power-broadened intensity PL,
even for low temperatures (4~K) and rather small cavity-dot coupling rates  ($g=20~\mu$eV, cf.~$\kappa=50~\mu$eV). 

\subsection{Power Broadening through Coherent Exciton Pumping and Cavity Emission}
\label{C:Pumped}


\begin{figure}[t!]
\centering\includegraphics[width=1\columnwidth]{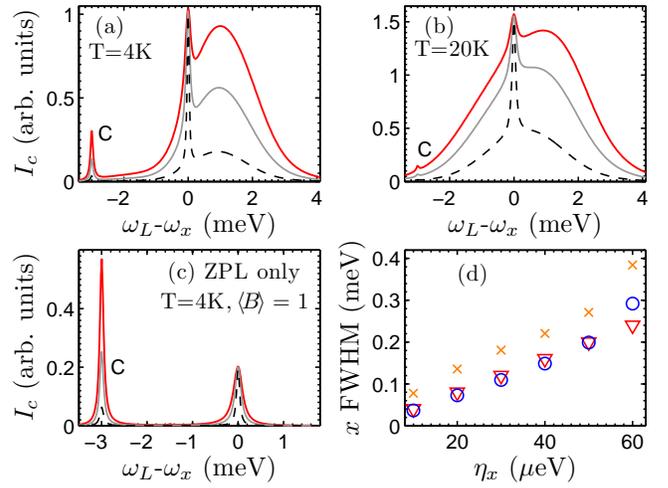}
\caption{(Color online) (a-b) Relative cavity mode intensity as a function of QD-laser detuning for a dot driven system at two different temperatures, (a) $T=4$~K and (b) $T=20$~K,
and for different values of the cw  Rabi frequency
($\eta_{x}=20$~$\mu$eV: black dashed line, $\eta_{x}=40$~$\mu$eV: grey, middle solid line,
 and $\eta_{x}=60$~$\mu$eV: red, upper
solid line). The cw field drives the QD which is detuned to the right of the cavity mode by $3$~meV 
($\Delta_{cx}=-3~$meV). 
Note that $\langle B\rangle=0.91$ at $T=4$~K and $\langle B\rangle=0.73$ at $T=20$~K. (c) Normalized cavity mode intensity in the presence of only ZPL broadening with $\langle B\rangle=1$ at $T=4$~K. (d) Plot of the $x$ FWHM of the intensity PL
at the QD resonance as a function of $\eta_c$. The orange crosses show the
 FWHM at $T=20$~K, the blue circles at $T=4$~K, and the red inverted
 triangles show the intensity FWHM with only ZPL broadening and $\langle B\rangle=1$ (no phonons).}
\label{fig:10n}
\end{figure}

\begin{figure}[t!]
\centering\includegraphics[width=1\columnwidth]{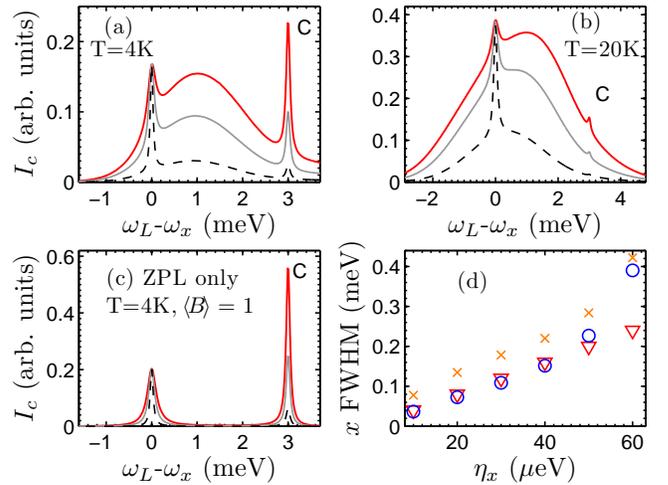}
\caption{(Color online)
As in Fig.~\ref{fig:10n} but with $\Delta_{cx}=3~$meV [cf.~$\Delta_{cx}=-3~$meV in Fig.~\ref{fig:10n}].}
\label{fig:11n}
\end{figure}

Next we study the cavity-emitted PL for different input powers of an exciton pump. In Figs.~\ref{fig:10n}(a-b) we plot the relative cavity mode intensity ($I_c$) as a function of exciton-laser detuning for a QD 
driven system at two temperatures, (a) $T=4$~K and (b) $T=20$~K, for various values of $\eta_x$. The cw field drives the QD
which is now detuned to the right (higher energy) of the cavity mode by $3$~meV. In Fig.~\ref{fig:10n}(c) we plot the relative
cavity mode intensity with only ZPL broadening and set $\langle B\rangle=1$ (i.e., no coherent or incoherent
coupling effects from phonons).
This then closely corresponds to an atomiclike power broadening model, but with the addition of pure dephasing of the ZPL---a model that
is commonly used to analyze semiconductor cavity-QED experiments. The power broadened intensity lineshape in the absence of  phonon
coupling 
can be explained
by considering two Lorentzian lineshapes centered at the two resonances, the relative oscillator strengths of which are 
qualitatively determined by their corresponding broadenings.

Without phonon coupling, as shown in Fig.~\ref{fig:10n}(c), an increasing cw drive causes power broadening of the QD exciton which decreases the oscillator strength relative to the cavity mode and also excites the cavity resonance more.
With phonon scattering processes [see Fig.~\ref{fig:10n}(a)],
the first major difference  we notice
 is the apparent narrowing of the QD resonance compared to Fig.~\ref{fig:10n}(c) which is due to the coherent renormalization of the Rabi frequency---this manifests in an effective drive whose magnitude decreases with increasing temperature; in addition, the phonon interactions reduce the dot-cavity coupling 
 through  $g\rightarrow \braket{B}g$.
Incoherent phonon coupling also introduces significant additional broadening of the QD exciton due to
the ${\cal L}(\sigma^+)$  process (see Table~\ref{tab:example1}), eventually resulting in a new
peak near the phonon cut off frequency (i.e., at the peak of the spectral bath
function for phonons, $\omega_b=1~$meV). The mean cavity photon numbers 
(and thus $I_c$) increase in the presence of phonons due to phonon-assisted processes whose magnitude also
increases with temperature (see Fig.~\ref{fig:10n}(c), which also has a larger ZPL).
Comparing the cavity PL characteristics in Fig.~\ref{fig:10n}(a-c), we see that
electron-phonon coupling plays 
a significant role in determining the intensity PL, with features that are not at all
explained with simple atomiclike MEs.


In Fig.~\ref{fig:10n}(d)  we plot the $x$ FWHM of $I_c$  at the QD resonance as a function of $\eta_x$, which is a typical measurement in experimental studies~\cite{ates3}.  The orange crosses show the intensity at $T=20$~K, the blue circles at $T=4$~K, and the inverted red triangles show the FWHM in the absence of any phonon coupling. In spite of
significant {\em reduction} of the effective Rabi frequency due to phonon coupling at high temperatures
(e.g., $\braket{B(4~K)}=0.91 \rightarrow \braket{B(20~K)}=0.77$), the $x$ FWHM at $T=20$~K is
{\em higher} than the FWHM calculated at $T=4$~K, which suggests that EID more than compensates for phonon-induced renormalization of the Rabi frequency.
It is noted from Table~\ref{tab:example1} earlier that increasing $\gamma$
 (and thus also $\Gamma^{\sigma^-}$, as they have the same Lindblad operators) reduces the mean cavity photon number. We also note from Table~\ref{tab:example4} that the primary contribution to the broadening are the 
 two scattering terms $\Gamma_{ph}^{\sigma^{+}}$ and $\Gamma_{ph}^{\sigma^{-}}$;
 these two 
  processes increase with temperature and driving strength and introduce additional
 broadening which reduces  $I_c$. At a pump rate of $\eta_x\sim 60~\mu$eV, we see a more rapid increase of the x FWHM with phonon coupling due to stronger cw laser-phonon coupling.
 For pump rates greater than
 $\eta_x \sim 40~\mu$eV, we also see a more rapid
 increase of the $x$ FWHM with phonon coupling.

In Figs.~\ref{fig:11n}(a-b) we plot the relative cavity-mode intensity PL for a dot driven system,
where  the exciton resonance is now detuned to the left of the cavity mode by $3$~meV.
  We again consider two different temperatures of the phonon reservoir, (a) $T=4$~K and (b) $T=20$~K,
  for various $\eta_x$. In Fig.~\ref{fig:11n}(c) we show the normalized cavity mode intensity with only ZPL broadening and $\langle B\rangle=1$. The generic features of Fig.~\ref{fig:11n} can be understood exactly along the lines of the arguments presented above for Fig.~\ref{fig:10n}.
 However, as discussed earlier,  here we obtain a significant reduction in the mean cavity photon number
  (less cavity feeding) since the cavity is now energetically higher which requires absorption of phonons---compare  Fig.~\ref{fig:10n}(a) and Fig.~\ref{fig:11n}(a). Moreover, the mean cavity photon number is smaller than that in the absence of phonons due to the renormalized 
  (reduced) dot-cavity coupling which, however, increases with temperature. In Fig.~\ref{fig:11n}(d) we plot the $x$ FWHM
  of the cavity emission 
  as a function of $\eta_x$ which shows very similar features to the $x$ FWHM data presented in Fig.~\ref{fig:10n}(d). We remark that the
   approximate $x$ FWHM values are extracted numerically by fitting with a simple Lorentzian model. As it is clear from the plots, the power broadening intensity PL in the presence of phonons are no longer represented by a simple system of coupled Lorentzians, and we observe
   pronounced non-Lorentzian lineshapes and clear signatures from the phonon bath spectral function. We  reiterate the point that there is a substantial discrepancy at large pumps between the $x$ FWHM  from an
effective Lindblad solution (EPME) and a full 
polaron ME  solution (see Table~\ref{tab:example4}). This highlights a breakdown of the EPME at large pumps which is not 
too surprising given the rather coarse approximations made  in its derivation. 
Finally, we comment that the $c$ FWHM (which can be extracted from the same PL data) does not show any 
power broadening, which is possibly 
also a result of the reasonably large QD-cavity detunings.


\subsection{Power Broadening through Coherent Cavity Pumping and Exciton Emission}

\begin{figure}[t!]
\centering\includegraphics[width=1\columnwidth]{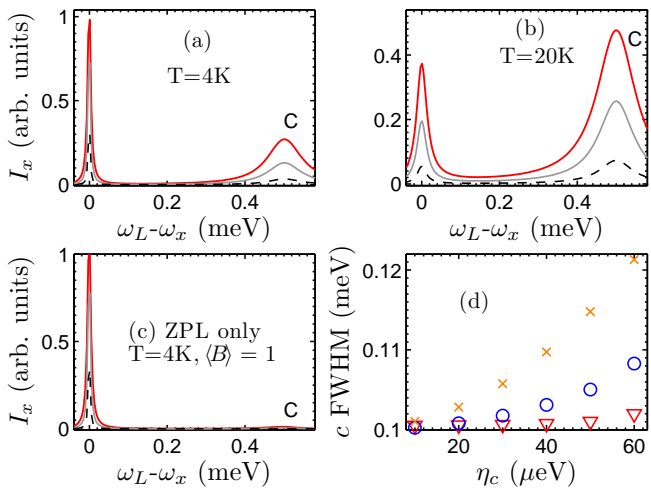}
\caption{(Color online)
As in Fig.~\ref{fig:10n}, but
for the exciton intensity obtained in the presence
of a cavity pump. The cw drive strengths
are $\eta_c=10,40, 60~\mu$eV.
In (d), the orange crosses show the
$c$ FWHM at $T=20$~K, the blue circles at $T=4$~K, and the inverted red triangles show the FWHM with only ZPL broadening and $\langle B\rangle=1$. The cavity-exciton detuning is
$\Delta_{cx}=0.5~$meV.
}
\label{fig:12n}
\end{figure}

In this subsection,
we focus on a cavity-excited system, where the emitted excited intensity ($I_x$) is detected through the 
 QD  emission, e.g., through (non-cavity mode) radiation modes. In Figs.~\ref{fig:12n}(a) and \ref{fig:12n}(b) we plot the relative exciton intensity, $I_{x}$, as a function of QD-laser detuning for a cavity driven system at two different temperatures, (a) $T=4$~K and (b) $T=20$~K, for various values of $\eta_c$. The cw field  drives the cavity mode which is  now detuned to the right of the QD
exciton by $0.5$~meV. In Fig.~\ref{fig:12n}(c) we show $I_x$ with no phonon
interactions and $\langle B\rangle=1$ (i.e., no coherent or incoherent effects from phonons, apart from
pure dephasing). 
We discern that the power broadened intensity PL in the absence of phonons is quite distinct from the case with finite phonon coupling.
On the one hand, we have lost the resonance at the phonon spectral function since
there is no longer a term that involved incoherent excitation through the phonon reservoir ($\Gamma_{ph}^{\sigma^{+}}$ process).
On the other hand, we  see a clear influence from phonon-induced exciton cavity feeding.
In particular, without phonon coupling we observe very
 little emission at the cavity mode frequency which further demonstrates that phonons play a significant role in
the power broadened PL lineshape through dot-cavity coupling (via cavity feeding). Also note that the broadening of 
the QD resonance increases due to enhanced dot-cavity coupling in the presence of phonons---which increases with temperature. However at higher temperatures the mean exciton number is decreased which is mainly due to phonon-induced reduction in $g\rightarrow \braket{B}\!g$.

In Fig.~\ref{fig:12n}(d) we plot the $c$ FWHM
 obtained via exciton emission ($I_x$), as a function of $\eta_c$. The $c$ FWHM in the presence of phonons at low temperatures ($T=4$~K) is very similar to the case with only ZPL broadening primarily because of the absence of EID, as the system is now cavity-driven, though 
 $g\rightarrow \braket{B}g$ is still temperature-dependent. However, the $c$ FWHM increases with temperature even though
$\braket{B}\!g$ reduces. We also note that the emission at the QD frequency is suppressed with increasing temperatures. With further increase in temperatures (i.e., above 20~K---not shown), the intensity PL is dominated by emission at the cavity mode.
Similar to the exciton pumped system, for pump rates
 of $\eta_c \sim 40~\mu$eV, we observe a more rapid
 increase of the $c$ FWHM with phonon coupling. For cavity excitation,
the effective Lindblad solution shown earlier provides a very good match for the $c$ FWHM for all 
$\eta_c$ pumps values studied (see Table~\ref{tab:example6}). This is primarily because the approximations made for a cavity driven system are less restrictive. For instance, there are only two Lindblad terms which is unlike the case of a QD-driven system where four Lindblad terms are needed (two of which
are $\eta_x$ dependent).

\begin{figure}[t!]
\centering\includegraphics[width=1\columnwidth]{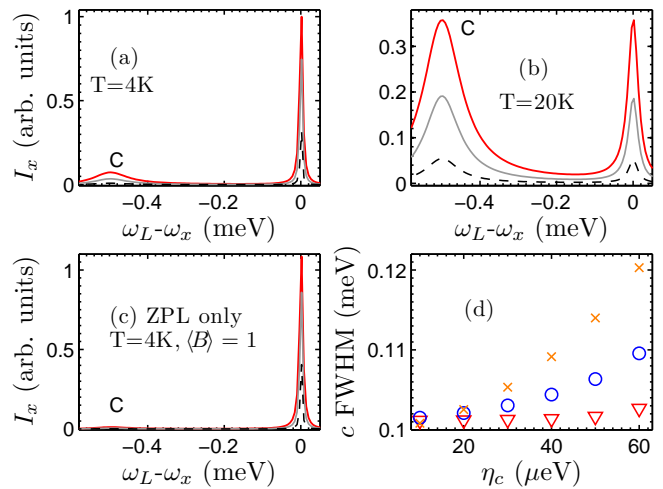}
\caption{(Color online)
As in Fig.~\ref{fig:11n}, but
for $\Delta_{cx}=-0.5~$meV.}
%
\label{fig:13n}
\end{figure}

In Figs.~\ref{fig:13n}(a) and \ref{fig:13n}(b) we again plot the relative exciton intensity ($I_x$) as a function of QD-laser detuning for a cavity driven system, but now with the cavity mode detuned to the left (lower energy) of the QD exciton  by $0.5$~meV.
 In Fig.~~\ref{fig:13n}(c) we show the normalized cavity mode intensity  with only ZPL broadening of the exciton and $\langle B\rangle=1$.
 We highlight the
 significant difference in the QD emission at the cavity mode energy for the two detunings at low temperature [Fig.~\ref{fig:12n}(a) and Fig.~\ref{fig:13n}(a)]---the cavity mode resonance is more pronounced with a positive $\Delta_{cx}$. This
  difference is due to the asymmetry in the phonon absorption and emission probability;
 at higher temperatures, however, the lineshapes becomes more similar.
  In Fig.~\ref{fig:13n}(d) we plot the $c$ FWHM 
  obtained through exciton emission
   as a function of the cw laser Rabi frequency ($\eta_c$), which is very similar to that in Fig.~\ref{fig:12n}(d). For these studies, the $c$ FWHM is mostly independent of the
  detuning of the dot-cavity system.
We also mention that $x$ FWHM (which can be extracted from the same PL data) only shows marginal power broadening ,
though 
it may increase with reduced QD-cavity detuning.

\subsection{Integrated Photoluminescence (IPL)}

\begin{figure}[h!]
\centering\includegraphics[width=1\columnwidth]{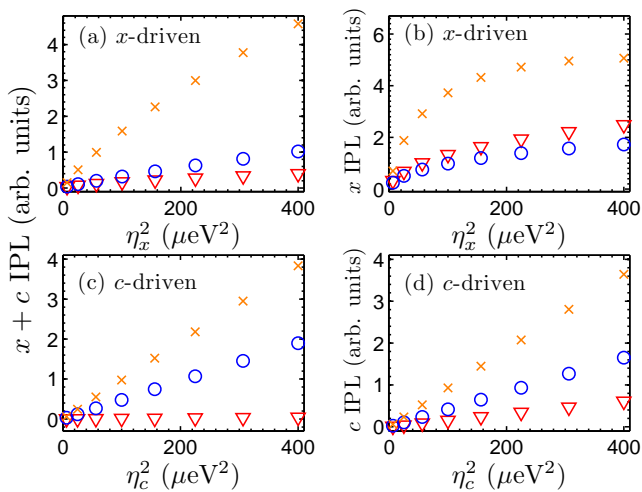}
\vspace{-0.5cm}
\caption{(Color online)  Integrated PL for a dot-driven system, (a) and (b), as measured through  cavity
emission, and the IPL of a cavity-driven system, (c) and (d), as measured via QD emission; both as a function of square of the cw drive for two different temperatures of
the phonon reservoir ($T=4$~K: blue circles , $T=20$~K: orange crosses); also plotted are the IPL lineshape in the presence of only ZPL broadening and $\langle B\rangle=1$ (inverted red triangles); the corresponding $\gamma'=2~\mu$eV (at 4~K). (a) The IPL of the {\em total} cavity and QD intensity of a dot-driven system as measured through cavity emission;  $\Delta_{cx}=3$~meV and the intensity PL are shown in Fig.~\ref{fig:11n}. (b)  IPL of the cavity mode intensity over the Lorentzian centered at the QD resonance frequency. (c) IPL of the QD exciton and cavity intensity of a cavity-driven system as measured through the QD; here $\Delta_{cx}=0.5$~meV and the intensity lineshapes are shown in Fig.~\ref{fig:13n}. (d) The IPL of the QD mode intensity over the Lorentzian centered at the cavity resonance frequency.}
\label{fig:14n}
\end{figure}

Finally, we study the  integrated photoluminescence  (IPL). In Figs.~\ref{fig:14n}(a) and \ref{fig:14n}(b) we plot the IPL lineshape for a dot-driven system as
 measured through the cavity, for the positively detuned cavity mode ($\Delta_{cx}=3$~meV). The total IPL of the cavity and exciton intensity of a dot-driven system plotted in Fig.~14(a) does not saturate with increasing drives---unlike the IPL over the Lorentzian centered at the exciton resonance frequency plotted in Fig. 14(b) (i.e., the integrated x FWHM, emitted via the cavity mode), which shows clear saturation. The Lorentzian lineshape at the QD exciton resonance is computed by subtracting  off the background phonon sidebands from the intensity PL. The lack of saturation of the IPL of the total cavity and QD intensity is attributed to the increasing excitation of the background phonon continuum with increasing cw drives which is also enhanced at higher temperatures.
%
These  theoretical trends 
are consistent with  the experimental results 
of Ates {\em et al.}~\cite{ates3}.

 In Figs.~\ref{fig:14n}(c) and \ref{fig:14n}(d) we show the IPL of the cavity-driven system (with $\Delta_{cx}=0.5$~meV) as measured via the QD exciton emission. Neither the total IPL of the cavity {\em and} exciton intensity of a cavity-driven system plotted in Fig. 14(c), nor the IPL over the Lorentzian centered only at the
cavity resonance frequency plotted in Fig. 14(d) (i.e., the integrated c FWHM, emitted via the QD) show any saturation. Furthermore, significant power broadening is found which increases with temperature of the phonon bath because of enhanced phonon-mediated coupling between the dot and the cavity mode.

\section{Conclusions}
\label{conclusions}

We have presented a detailed analysis of the intensity PL power broadening of a semiconductor cavity-QED system
under coherent excitation of either the QD exciton or the cavity mode. In particular, we have included the interaction of the QD with an acoustic phonon environment at a microscopic level, while also accounting for exciton-cavity coupling in the regime
of cavity-QED.
 We utilized a time-convolutionless ME approach in the polaron frame to study the cavity-QED dynamics. The interaction of the phonon reservoir with the QD is nonperturbative and is limited mainly by the validity of the IBM (and the single exciton picture) at high temperatures. This approach  enables us to treat the coherent interaction between the QD, cavity and the cw laser field to all orders. Central to this approach is the need to account for the internal coupling effects which preserves the detail balanced condition on the system density operator.
 Our theory also points out some major flaws and restrictions of atomiclike MEs for modeling
 coherent excitation regimes in  semiconductor QD-cavity systems.

Using material parameters close to those measured in 
related semiconductor experiments,
various intensity PL lineshapes were studied as a function of the excitation pump rate, for different
  temperatures of the phonon bath. We computed the
full intensity PL lineshape over a range of frequencies and extracted approximate Lorentzianlike linewidths of the  $c/x$ FWHM of the power broadening intensity PL for both the QD-driven and cavity-driven system. The interaction of the QD with the phonon reservoir is seen to introduce qualitatively different features in the intensity PL, especially at high temperatures, and is quite distinct from the case of a typical power broadening lineshape of a two-level atom. We  found a significant narrowing of the QD resonance due to the
 coherent renormalization of the Rabi frequency resulting in a reduced effective drive. The interaction with the phonon reservoir also reduced the dot-cavity coupling due to renormalization of $g\rightarrow \braket{B}g$. Phonon coupling further introduces additional broadening of the QD exciton due to EID and results
in highly  non-Lorentzian lineshapes for the intensity PL; via incoherent excitation from the phonon bath, clear signatures of the
phonon bath spectral function appear in the exciton driven PL.
For a cavity driven system,  the mean cavity photon number (which is proportional to the cavity mode
 intensity), was also found to change in the presence of phonon-assisted processes which depended 
 sensitively on dot-cavity detuning and the  temperature of the phonon bath.

To help  explain the underlying physics of electron-phonon scattering in these
cavity-QED systems,
we  derived an effective phonon ME (EPME) of the Lindblad form, which facilitates a very simple numerical 
solution to the full ME and
also allows the extraction of various phonon-induced scattering rates in a physically
meaningfully way. In particular,  we identified specific phonon-mediated processes which cause EID of the QD 
exciton resonance
and incoherent exciton pumping. We also identified Lindblad superoperators which mediate coherent interactions between the QD and the cavity mode, resulting in exciton-cavity feeding---a phenomenon that is becoming more familiar
in semiconductor cavity-QED (e.g., see Refs.~\onlinecite{JiaoJPC2008,SufczynskiPRL:2009,ota,ates2,Dalacu:PRB2010,HohenesterPRB:2009,HohenesterPRB:2010,hughes2,SavonaPRB:2010, roy_hughes,Calic:PRL11}). We also studied the regimes of validity of the effective Lindblad solution and found
that, for relatively large QD-cavity detunings, 
the effective Lindblad solution produces a very good fit to the full polaron ME solution. However, discrepancies can occur
for exciton pumped systems with increasing drives, where one requires the full
polaron ME to obtain a more accurate
FWHM of the exciton resonance.
Using these polaronic ME formalisms, we found that phonon-induced EID and incoherent coupling between the QD and the cavity is fundamental
to obtain a complete picture of power broadening in these semiconductor systems. 
In particular, our results demonstrate that the cavity
emitted
QD-driven intensity PL can  display quite profound signatures of the phonon spectral bath
function.
A cavity driven system also contains clear signatures of the 
phonon bath, containing power broadened features that are significantly different
to those obtained in atomic QED. We have  demonstrated that pure dephasing of the ZPL, frequently cited
as being responsible for phonon-scattering effects such as cavity feeding, actually only plays a very minor role here;
the physics of the phonon scattering processes that we identity are manifestly 
different to the
physics of coupled Lorentzian oscillators, either with or without
pure dephasing of the ZPL. 

The results presented in this paper should  be broad interest for interpreting current and
 future experimental data from semiconductor cavity-QED systems. A more rigorous analysis would use the full polaron ME and
  a less rigorous approach
  would use the
   EPME; the latter approach allows for a simple and intuitive picture of the underlying physics
   and can be accurate in certain excitation regimes. 
   Our power broadening results were chosen to highlight new
 regimes where phonon effects are more readily visible and can serve to direct experimental focus in 
 those regimes.
 For this work, we have also focused our studies to the domain of cw excitation, but
 the extension to pulsed systems is straightforward and will be presented
 elsewhere.

\section*{Acknowledgments}
This work was supported by the National Sciences and Engineering Research Council
of Canada and the Canadian Foundation for Innovation.
We thank H.~J. Carmichael for useful discussions.

\begin{widetext}
\appendix
\section{Effective Lindblad master equations with simplified electron-phonon coupling}
\label{appendix1}
In this appendix we provide some technical details into the derivation of our effective Lindblad master equation, Eq.~(\ref{effectiveME}).
We exemplify for the relatively simple case of a cavity driven system, where $\hat{X}_{g}= \hbar g(\hat{a}^{\dagger}\hat{\sigma}^{-}+\hat{\sigma}^{+}\hat{a})$ and $\hat{X}_{u}= i\hbar g(\hat{\sigma}^{+}\hat{a}-\hat{a}^{\dagger}\hat{\sigma}^{-})$. The case of a dot driven system can also be derived using similar steps but with a few additional approximations discussed at the end of this appendix. The integrand inside the  phonon integral in Eq.~(\ref{sec3eq3}), $\int^{\infty}_{0}d\tau\sum_{m=g,u}(G_{m}(\tau)[\hat{X}_{m},e^{-iH_{sys}^{\prime}\tau/\hbar}\hat{X}_{m}e^{iH_{sys}^{\prime}\tau/\hbar}\rho(t)]+H.c.)$, can be approximated (e.g., for $m=g$) as follows:
\begin{align}
\label{append1}
& G_{g}(\tau)[\hat{X}_{g},e^{-iH_{sys}^{\prime}\tau/\hbar}\hat{X}_{g}e^{iH_{sys}^{\prime}\tau/\hbar}\rho(t)]+H.c. \nonumber \\
& \simeq \hbar^{2}g^{2}G_{g}(\tau)(\hat{a}^{\dagger}\hat{\sigma}^{-}+\hat{\sigma}^{+}\hat{a})(\hat{a}^{\dagger}\hat{\sigma}^{-}
e^{-i\Delta_{cx} t}
+\hat{\sigma}^{+}\hat{a}e^{i\Delta_{cx} t})\rho(t) - \hbar^{2}g^{2}G_{g}(\tau)(\hat{a}^{\dagger}\hat{\sigma}^{-}
e^{-i\Delta_{cx} t}+\hat{\sigma}^{+}\hat{a}e^{i\Delta_{cx} t})\rho(t)(\hat{a}^{\dagger}\hat{\sigma}^{-}+\hat{\sigma}^{+}\hat{a}) \nonumber \\
& +\hbar^{2}g^{2}G^{*}_{g}(\tau)\rho(t)(\hat{a}^{\dagger}\hat{\sigma}^{-}e^{-i\Delta_{cx} t}+\hat{\sigma}^{+}\hat{a}e^{i\Delta_{cx} t})(\hat{a}^{\dagger}\hat{\sigma}^{-}+\hat{\sigma}^{+}\hat{a}) -\hbar^{2}g^{2}G^{*}_{g}(\tau)(\hat{a}^{\dagger}\hat{\sigma}^{-}+\hat{\sigma}^{+}
\hat{a})\rho(t)(\hat{a}^{\dagger}\hat{\sigma}^{-}e^{-i\Delta_{cx} t}+\hat{\sigma}^{+}
\hat{a}e^{i\Delta_{cx} t}), \nonumber \\
\end{align}
where we have used
\begin{equation}
e^{-iH_{sys}^{\prime}\tau/\hbar}\hat{X}_{m}e^{iH_{sys}^{\prime}\tau/\hbar}\simeq e^{-iH_{0}^{\prime}\tau/\hbar}\hat{X}_{m}e^{iH_{0}^{\prime}\tau/\hbar},
\end{equation}
with $H^{\prime}_{0} = \hbar\Delta_{xL}\hat{\sigma}^{+}\hat{\sigma}^{-}+\hbar\Delta_{cL}\hat{a}^{\dagger}\hat{a}$. This corresponds to approximating $H^{\prime}_{\rm sys}$ with $H^{\prime}_{0}$ in the exponential phase and is expected to be only valid when the dot-cavity detuning is large compared to $g$. It follows that
\begin{eqnarray}
e^{-iH_{0}^{\prime}\tau/\hbar}\hat{a}^{\dagger}\hat{\sigma}^{-}e^{iH_{0}^{\prime}\tau/\hbar}&=&e^{-i\Delta_{cx} t}\hat{a}^{\dagger}\hat{\sigma}^{-},  \\
e^{-iH_{0}^{\prime}\tau/\hbar}\hat{\sigma}^{+}\hat{a}e^{iH_{0}^{\prime}\tau/\hbar}&=&e^{i\Delta_{cx} t}\hat{\sigma}^{+}\hat{a}.
\end{eqnarray}
Consequently, this allows us to write an effective Lindblad ME that can (e.g., see Figs~\ref{fig:3n}-\ref{fig:5n}) 
reproduce the full polaron ME solution over a   range of dot-laser and cavity-exciton detunings.
The major advantage of this approach (used on its own or as a compliment) is that it is simpler than the
full polaron ME approach and allows one to extract various, physically meaningful, scattering 
 processes associated with electron-phonon interactions. Equation~(\ref{append1}) can subsequently be rewritten as
\begin{align}
\label{append3}
G_{g}(\tau)[\hat{X}_{g},e^{-iH_{sys}^{\prime}\tau/\hbar}\hat{X}_{g}e^{iH_{sys}^{\prime}\tau/\hbar}\rho(t)]+H.c.
&\simeq \hbar^{2}g^{2}\left [G_{g}(\tau)e^{-i\Delta_{cx} t}\hat{\sigma}^{+}\hat{a}\hat{a}^{\dagger}\hat{\sigma}^{-}\rho(t)+G_{g}(\tau)e^{i\Delta_{cx} t}\hat{a}^{\dagger}\hat{\sigma}^{-}\hat{\sigma}^{+}\hat{a}\rho(t)\right ] \ \ \ \ \ \ \ \nonumber \\
&-\hbar^{2}g^{2} \left [ G_{g}(\tau)e^{-i\Delta_{cx} t}\hat{a}^{\dagger}\hat{\sigma}^{-}\rho(t)\hat{\sigma}^{+}\hat{a}+G_{g}(\tau)e^{i\Delta_{cx} t}\hat{\sigma}^{+}\hat{a}\rho(t)\hat{a}^{\dagger}\hat{\sigma}^{-} \right ] \nonumber \\
&+\hbar^{2}g^{2} \left [ G^{*}_{g}(\tau)e^{-i\Delta_{cx} t}\rho(t)\hat{a}^{\dagger}\hat{\sigma}^{-}\hat{\sigma}^{+}\hat{a}+G^{*}_{g}(\tau)e^{i\Delta_{cx} t}\rho(t)\hat{\sigma}^{+}\hat{a}\hat{a}^{\dagger}\hat{\sigma}^{-} \right ] \nonumber \\
&-\hbar^{2}g^{2} \left [ G^{*}_{g}(\tau)e^{-i\Delta_{cx} t}\hat{\sigma}^{+}\hat{a}\rho(t)\hat{a}^{\dagger}\hat{\sigma}^{-}+G^{*}_{g}(\tau)e^{i\Delta_{cx} t}\hat{a}^{\dagger}\hat{\sigma}^{-}\rho(t)\hat{\sigma}^{+}\hat{a} \right ] \, ,
\end{align}
where we have neglected the other terms (such as $\hat{\sigma}^{+}\hat{a}\hat{\sigma}^{+}\hat{a}\rho$, $\hat{\sigma}^{+}\hat{a}\rho\hat{\sigma}^{+}\hat{a}$ and so on) as they do not contribute to the ME. Equation~(\ref{append3}) can now be simplified to
\begin{align}
\label{append4}
G_{g}(\tau)[\hat{X}_{g},e^{-iH_{sys}^{\prime}\tau/\hbar}\hat{X}_{g}e^{iH_{sys}^{\prime}\tau/\hbar}\rho(t)]+H.c.
&\simeq \hbar^{2} g^{2}\,{\rm Re}[G_{g}(\tau)e^{-i\Delta_{cx} t}](\hat{\sigma}^{+}\hat{a}\hat{a}^{\dagger}\hat{\sigma}^{-}\rho(t)
+\rho(t)\hat{\sigma}^{+}\hat{a}\hat{a}^{\dagger}\hat{\sigma}^{-}-2\hat{a}^{\dagger}\hat{\sigma}^{-}
\rho(t)\hat{\sigma}^{+}\hat{a}) \nonumber \\
&+\hbar^{2}g^{2}\,{\rm Re}[G_{g}(\tau)e^{i\Delta_{cx} t}](\hat{a}^{\dagger}\hat{\sigma}^{-}\hat{\sigma}^{+}\hat{a}
\rho(t)
+\rho(t)\hat{a}^{\dagger}\hat{\sigma}^{-}\hat{\sigma}^{+}\hat{a}
-2\hat{\sigma}^{+}\hat{a}\rho(t)\hat{a}^{\dagger}\hat{\sigma}^{-}) \nonumber \\
&+i\hbar^{2}g^{2}\,{\rm Im}[G_{g}(\tau)e^{-i\Delta_{cx} t}](\hat{\sigma}^{+}\hat{a}\hat{a}^{\dagger}\hat{\sigma}^{-}
\rho(t)-\rho(t)\hat{\sigma}^{+}\hat{a}\hat{a}^{\dagger}\hat{\sigma}^{-})
\nonumber \\
&+i\hbar^{2}g^{2}\,{\rm Im}[G_{g}(\tau)e^{i\Delta_{cx} t}](\hat{a}^{\dagger}\hat{\sigma}^{-}\hat{\sigma}^{+}\hat{a}\rho(t)-\rho(t)\hat{a}^{\dagger}
\hat{\sigma}^{-}\hat{\sigma}^{+}\hat{a}).
\end{align}
Using
$\sum_{m=g,u}G_{m}(\tau)=\langle B\rangle^2 (e^{\phi(\tau)}-1)$,
 the defined scattering rates become,
\begin{align}
\Gamma_{ph}^{\sigma^{+}a/a^{\dagger}\sigma^{-}}\!=2\langle B\rangle^2 g^{2}\,{\rm Re}\left [\int_{0}^{\infty}d\tau
e^{\pm i\Delta_{cx} \tau}\left (e^{\phi(\tau)} -1 \right ) \right],
\end{align}
while the frequency-shifts,
\begin{align}
  \Delta_{ph}^{\sigma^{+}a/a^{\dagger}\sigma^{-}}\!=\langle B\rangle^2 g^{2}\,{\rm Im}\left [\int_{0}^{\infty}
  d\tau e^{\pm i\Delta_{cx} \tau}\left (e^{\phi(\tau)}-1\right ) \right],
\end{align}
   which are used to obtain the contribution of the phonon integral in the Lindblad form, $L_{\rm ph}(\rho)$, and
    the effective Hamiltonian defined in Eq.~(\ref{sec3eqfinal3}).

For the case of a dot-driven system, we have an additional complication due to the coherent term driving
 through the  exciton-phonon bath. Using similar steps as above, and also neglecting  contributions involving cross terms between operators $\hat{\sigma}^{+}$ ($\hat{\sigma}^{-}$) 
 and $\hat{\sigma}^{+}\hat{a}$ ($\hat{a}^{\dagger}\hat{\sigma}^{-}$),
  which scale as $g\eta_{x}$, we again obtain an effective Lindblad form of the ME. We evaluate the exponential phase terms involving the full system Hamiltonian by replacing $H^{\prime}_{sys}$ with $H^{\prime}_{0}$.


\end{widetext}




\begin{thebibliography}{99}


\bibitem{hohenester}U. Hohenester, {\it Optical properties of semiconductor nanostructures: Decoherence versus Quantum Control},
Handbook of Theoretical and Computational Nanotechnology (2006)

\bibitem{entangled1} N. Akopian, N. H. Lindner, E. Poem, Y. Berlatzky, J. Avron, D. Gershoni, B. D. Gerardot, and P. M. Petroff,
{\it Entangled Photon Pairs from Semiconductor Quantum Dots}, Phys. Rev. Lett. {\bf 96}, 130501 (2006).

\bibitem{entangled2} A. Muller, W. Fang, J. Lawall, and G. S. Solomon,
{\it Creating Polarization-Entangled Photon Pairs from a Semiconductor Quantum Dot Using the Optical Stark Effect}, Phys. Rev. Lett. {\bf 103}, 217402 (2009).

\bibitem{entangled3} R. B. Patel, A. J. Bennett, K. Cooper, P. Atkinson, C. A. Nicoll, D. A. Ritchie, and A. J. Shields,
{\it Postselective Two-Photon Interference from a Continuous Nonclassical Stream of Photons Emitted by a Quantum Dot}, Phys. Rev. Lett. {\bf 100}, 207405 (2008).

\bibitem{ates1}S. Ates, S. M. Ulrich, S. Reitzenstein, A. L\"offler, A. Forchel, and P. Michler,
{\it Post-Selected Indistinguishable Photons from the Resonance Fluorescence of a Single Quantum Dot in a Microcavity}, Phys. Rev. Lett. {\bf 103}, 167402 (2009).

\bibitem{santori}C. Santori, D. Fattal, J. Vu\ifmmode \check{c}\else \v{c}\fi{}kovi\ifmmode \acute{c}\else \'{c}\fi{}, G. S. Solomon, and Y. Yamamoto,
{\it Indistinguishable photons from a single-photon device}, Nature {\bf 419}, 594 (2002).

\bibitem{SC:StrongCoupling1}
See, e.g.,
J. P. Reithmaier, G. Seogonk, A. L\"offler, C. Hofmann, S. Kuhn, S. Reitzenstein,
L. V. Keldysh, V. D. Kulakovskii, T. L. Reinecke, and A. Forchel,
{\it Strong coupling in a single quantum dot semiconductor microcavity system}, Nature  {\bf 432}, 197 (2004).

\bibitem{SC:StrongCoupling2}
T. Yoshie, A. Scherer, J. Hendrickson, G. Khitrova, H. M. Gibbs, G. Rupper, C. Ell, O. B. Shchekin, and D. G.
Deppe, {\it Vacuum Rabi splitting with a single quantum dot in a photonic crystal nanocavity}, Nature \textbf{432}, 200 (2004).


\bibitem{press}D. Press, S. G\"otzinger, S. Reitzenstein, C. Hofmann, A. L\"offler, M. Kamp, A. Forchel, and Y. Yamamoto,
{\it Photon Antibunching from a Single Quantum-Dot-Microcavity System in the Strong Coupling Regime}, Phys. Rev. Lett. {\bf98}, 117402 (2007).

\bibitem{muller}A. Muller, E. B. Flagg, P. Bianucci, X. Y. Wang, D. G. Deppe, W. Ma, J. Zhang, G. J. Salamo, M. Xiao, and C. K. Shih,
{\it Resonance Fluorescence from a Coherently Driven Semiconductor Quantum Dot in a Cavity}, Phys. Rev. Lett. {\bf99}, 187402 (2007).

\bibitem{flagg}E. B. Flagg, A. Muller, J. W. Robertson, S. Founta, D. G. Deppe, M. Xiao, W. Ma, G. J. Salamo,  and C. K. Shih,
{\it Resonantly driven coherent oscillations in a solid-state quantum emitter}, Nature Physics {\bf 5}, 203 (2009).

\bibitem{vamivakas} A. N. Vamivakas, Y. Zhao, C.-Y. Lu, and  M. Atat\"ure,
{\it Spin-resolved quantum-dot resonance fluorescence}, Nature Physics {\bf 5}, 198 (2009).


\bibitem{ates2}S. Ates, S. M. Ulrich, A. Ulhaq, S. Reitzenstein, A. L\"offler, S. H\"ofling, A. Forchel,  and P. Michler,
{\it Non-resonant dot-cavity coupling and its potential for resonant single-quantum-dot spectroscopy},
Nature Photonics {\bf3}, 724 (2009).

\bibitem{jelena_arka}
A. Majumdar, E. D. Kim, Y. Gong, M. Bajcsy, 
 and J. Vu\ifmmode \check{c}\else \v{c}\fi{}kovi\ifmmode \acute{c}\else \'{c}\fi, Phonon mediated off-resonant quantum dot�cavity coupling under resonant excitation of the quantum dot, Phys. Rev. B 84, 085309 (2011); see also
A. Majumdar, E. D. Kim, Y. Gong, M. Bajcsy, P. Petroff, and J. Vu\ifmmode \check{c}\else \v{c}\fi{}kovi\ifmmode \acute{c}\else \'{c}\fi{},
{\it Probing of single quantum dot dressed states via an off-resonant cavity}, Phys. Rev. B, {\bf 84}, 085310 (2011).



\bibitem{HennessyNature:2007}
K. Hennessy, A. Badolato, M. Winger, A. Atat\"ure, S. Falt, E. L. Hu, and A.~Imamogl{\u u},
{\it Quantum nature of a strongly coupled single quantum dot-cavity system}, Nature {\bf 445}, 896 (2007).

\bibitem{KaniberPRB:2008}
M. Kaniber, A. Laucht, A. Neumann, J.M. Villas-B\^oas, M. Bichler, M.-C. Amann, and J. J. Finley,
{\it Investigation of the nonresonant dot-cavity coupling in two-dimensional photonic crystal nanocavities}, Phys. Rev. B {\bf 77},
161303(R) (2008).


\bibitem{RuthOE:2007}
R. Oulton, B.D. Jones, S. Lam, A.R.A. Chalcraft, D. Szymanski, D. O'Brien, T.F. Krauss, D. Sanvitto, A. M. Fox, D.M. Whittaker, M. Hopkinson, and M.S. Skolnick,
{\it Polarized Quantum Dot Emission from Photonic Crystal Nanocavities studied under Mode Resonant Excitation}, {\bf 15} Opt. Express, 17221 (2007).


\bibitem{SufczynskiPRL:2009}
J. Suffczynski, A. Dousse, K. Gauthron, A. Lemaitre, I. Sagnes, L. Lanco, J. Bloch, P. Voisin, and P. Senellart,
{\it Origin of the optical emission within the cavity mode of coupled quantum dot-cavity systems}, Phys. Rev. Lett. {\bf 103}, 027401 (2009).

\bibitem{TawaraOE:2009}
T. Tawara, H. Kamada, S. Hughes, H. Okamoto, M. Notomi, and T. Sogawa,
{\it Cavity mode emission in weakly coupled quantum dot - cavity systems}, Opt. Express {\bf 17}, 6643 (2009).

\bibitem{ota}Y. Ota, S. Iwamoto, N. Kumagai, and Y. Arakawa,
{\it Impact of electron-phonon interactions on quantum-dot cavity quantum electrodynamics}, e-print: arXiv:0908.0788 (2009).

\bibitem{Dalacu:PRB2010}
D. Dalacu, K. Mnaymneh, V. Sazonova, P. J. Poole, G. C. Aers, J. Lapointe, R. Cheriton,
A. J. SpringThorpe, and R. L. Williams,
{\it
Deterministic emitter-cavity coupling using a single-site controlled quantum dot}, 
Phys Rev B {\bf 82}, 033301 (2010).

\bibitem{Calic:PRL11}
M. Calic, P. Gallo, M. Felici, K. A. Atlasov, B. Dwir, A. Rudra, G. Biasiol, L. Sorba, G. Tarel, V. Savona, and E. Kapon,
{\it Phonon-Mediated Coupling of InGaAs/GaAs Quantum-Dot Excitons to Photonic Crystal Cavities}, Phys. Rev. Lett. {\bf 106}, 227402 (2011).

\bibitem{hughes1}  F. Milde, A. Knorr, and S. Hughes,
{\it Role of electron-phonon scattering on the vacuum Rabi splitting of a single-quantum dot and a photonic crystal nanocavity}, Phys. Rev. B {\bf 78}, 035330 (2008).

\bibitem{hughes2} S. Hughes, P. Yao, F. Milde, A. Knorr, D. Dalacu, K. Mnaymneh, V. Sazonova, P. J. Poole, G. C. Aers, J. Lapointe, R. Cheriton,  and R. L. Williams,
{\it Influence of electron-acoustic phonon scattering on off-resonant cavity feeding within a strongly coupled quantum-dot cavity system}, Phys. Rev. B {\bf 83}, 165313 (2011).



\bibitem{JiaoJPC2008}J. Xue, K-D Zhu and H. Zheng,
{\it Detuning effect in quantum dynamics of a strongly coupled single quantum dot–cavity system}, J. Phys. C {\bf 20}, 325209 (2008).

\bibitem{HohenesterPRB:2009}
U. Hohenester, A. Laucht, M. Kaniber, N. Hauke, A. Neumann, A. Mohtashami, M. Selinger, M. Bichler, and J. J. Finley, 
{\it Phonon-assisted transitions from quantum dot excitons to cavity photons}, Phys. Rev. B {\bf 81}, 201311 (2009).

\bibitem{HohenesterPRB:2010}
 U. Hohenester,
{\it Cavity quantum electrodynamics with semiconductor quantum dots: Role of phonon-assisted cavity feeding}, Phys. Rev. B {\bf 81}, 155303 (2010).

\bibitem{kaer}P. Kaer, T. R. Nielsen, P. Lodahl, A.-P. Jauho, and J. M\o{}rk,
{\it Non-Markovian Model of Photon-Assisted Dephasing by Electron-Phonon Interactions in a Coupled Quantum-Dot--Cavity System}, Phys. Rev. Lett. {\bf 104}, 157401 (2010).

\bibitem{SavonaPRB:2010}
 G. Tarel and V. Savona,
{\it Linear spectrum of a quantum dot coupled to a nanocavity}, Phys. Rev. B {\bf 81}, 075305 (2010).

\bibitem{jelena1}A. Majumdar, A. Faraon, E. D. Kim, D. Englund, H. Kim, P. Petroff ,and J. Vu\ifmmode \check{c}\else \v{c}\fi{}kovi\ifmmode \acute{c}\else \'{c}\fi{},
{\it Linewidth broadening of a quantum dot coupled to an off-resonant cavity}, Phys. Rev. B, {\bf 82}, 045306 (2010).



\bibitem{auger} P. P. Paskov, P. O. Holtz, S. Wongmanerod, B. Monemar, J. M. Garcia, W. V. Schoenfeld, and P. M. Petroff,
{\it Auger processes in InAs self-assembled quantum dots}, Physica E {\bf 6}, 440 (2000).

\bibitem{auger1}M. Winger, T. Volz, G. Tarel, S. Portolan, A. Badolato, K. J. Hennessy, E. L. Hu, A. Beveratos, J. Finley, V. Savona, and
A. Imamo\ifmmode \breve{g}\else \u{g}\fi{}lu, 
{\it Explanation of Photon Correlations in the Far-Off-Resonance Optical Emission from a Quantum-Dot–Cavity System}, Phys. Rev. Lett. {\bf 103} 207403 (2009)

\bibitem{auger2} R. V\"olkl, M. Griesbeck, S. A. Tarasenko, D. Schuh, W. Wegscheider, C. Sch\"uller, and T. Korn, 
{\it Spin dephasing and photoinduced spin diffusion in a high-mobility two-dimensional electron system embedded in a GaAs-(Al,Ga)As quantum well grown in the [110] direction}, Phys. Rev. B, {\bf 83}, 241306 (2011)

\bibitem{auger3} A. Laucht, M. Kaniber, A. Mohtashami, N. Hauke, M. Bichler, and J. J. Finley, 
{\it Temporal monitoring of nonresonant feeding of semiconductor nanocavity modes by quantum dot multiexciton transitions}, Phys. Rev. B, {\bf 81}, 241302 (2010)

\bibitem{ates3}A. Ulhaq, S. Ates, S. Weiler, S. M. Ulrich, S. Reitzenstein, A. L\"offler, S. H\"ofling, L. Worschech, A. Forchel, and P. Michler,
{\it Linewidth broadening and emission saturation of a resonantly excited quantum dot monitored via an off-resonant cavity mode}, Phys. Rev. B {\bf 82}, 045307 (2010).






\bibitem{besombes}L. Besombes, K. Kheng, L. Marsal, and H. Mariette,
{\it Acoustic phonon broadening mechanism in single quantum dot emission}, Phys. Rev. B {\bf 63}, 155307 (2001).





\bibitem{Favero:PRB03} E. Peter, J. Hours, P. Senellart, A. Vasanelli, A. Cavanna, J. Bloch, and J. M. G\'erard, 
{\it Phonon sidebands in exciton and biexciton emission from single GaAs quantum dots}, Phys. Rev. B {\bf 69}, 041307 (2004).

\bibitem{Peter:PRB04}
I. Favero, G. Cassabois, R. Ferreira, D. Darson, C. Voisin, J. Tignon, C. Delalande, G. Bastard, Ph. Roussignol, and J. M. G\'erard, 
{\it Acoustic phonon sidebands in the emission line of single InAs/GaAs quantum dots}, Phys. Rev. B {\bf 68}, 233301 (2003).

\bibitem{stuttgart_prl}S. M. Ulrich, S. Ates, S. Reitzenstein, A. L\"offler, A. Forchel and P. Michler,
{\it Dephasing of Mollow Triplet Sideband Emission of a Resonantly Driven Quantum Dot in a Microcavity}, Phys. Rev. Lett. {\bf 106}, 247402 (2011).


\bibitem{roy_hughes}C. Roy and S. Hughes,
{\it Phonon-dressed Mollow triplet in the regime of cavity quantum electrodynamics: Excitation-induced dephasing and nonperturbative cavity feeding effects}, Phys. Rev. Lett. {\bf 106}, 247403 (2011).

\bibitem{nazir2}D. P. S. McCutcheon and A. Nazir,
{\it Quantum dot Rabi rotations beyond the weak exciton-phonon coupling regime}, New J. Phys. {\bf 12}, 113042 (2010).











\bibitem{nazir1}A. Nazir, {\it Photon statistics from a resonantly driven quantum dot}, Phys. Rev. B {\bf 78}, 153309 (2008).

\bibitem{mahan}G. D. Mahan, {\it Many-Particle Physics}, Plenum, New York, 1990.

\bibitem{krum}B. Krummheuer, V. M. Axt, and T. Kuhn, {\it Theory of pure dephasing and the resulting absorption line shape in semiconductor quantum dots}, Phys. Rev. B {\bf 65}, 195313 (2002).

\bibitem{imamoglu}I. Wilson-Rae  and A. Imamo\ifmmode \breve{g}\else \u{g}\fi{}lu,
{\it Quantum dot cavity-QED in the presence of strong electron-phonon interactions}, Phys. Rev. B {\bf65}, 235311 (2002).



\bibitem{wurger}A. W\"urger, {\it Strong-coupling theory for the spin-phonon model}, Phys. Rev. B {\bf 57}, 347 (1998).

 \bibitem{note3}
 We remark that, in fact, for our cw studies, the more complicated non-local ME yields the same result
 as our simpler time-local ME. We will show this directly in a future publication.

\bibitem{mogilevtsev1}D. Mogilevtsev, A. P. Nisovtsev, S. Kilin, S. B. Cavalcanti, H. S. Brandi, and L. E. Oliveira,
{\it Driving-Dependent Damping of Rabi Oscillations in Two-Level Semiconductor Systems}, Phys. Rev. Lett. {\bf 100}, 017401 (2008).


\bibitem{ramsay2}A. J. Ramsay, A. V. Gopal, E. M. Gauger, A. Nazir, B. W. Lovett, A. M. Fox, and M. S. Skolnick,
{\it Damping of Exciton Rabi Rotations by Acoustic Phonons in Optically Excited $InGaAs/GaAs$ Quantum Dots}, Phys. Rev. Lett. {\bf 104}, 017402 (2010).

\bibitem{carmichael}H. J. Carmichael and D. F. Walls,
{\it Master equation for strongly interacting systems}, J. Phys. A: Math. Nucl. Gen. {\bf 6}, 1552 (1973)


\bibitem{sajeev1}M. Florescu and S. John, {\it Single-atom switching in photonic crystals}, Phys. Rev. A {\bf 64}, 033801 (2001).


\bibitem{tanas1} A. Kowalewska-Kudlaszyk and R. Tanas,
{\it Generalized master equation for a two-level atom in a strong field and tailored reservoirs}, J. Mod. Opt. {\bf 48} 347 (2001).


\bibitem{BorriPRL:2001} P. Borri, W. Langbein, S. Schneider, U. Woggon, R. L. Sellin, D.
Ouyang, and D. Bimberg, {\it Ultralong Dephasing Time in InGaAs Quantum Dots}, Phys. Rev. Lett. {\bf 87}, 157401 (2001).


\bibitem{Rudin:PRL06}S. Rudin, T. L. Reinecke, and M. Bayer, {\it Temperature dependence of optical linewidth in single InAs quantum dots}, Phys. Rev. B {\bf 74}, 161305(R).
(2006).

\bibitem{Zimmermann:PRL04}
E. A. Muljarov and R. Zimmermann, {\it Dephasing in Quantum Dots: Quadratic Coupling to Acoustic Phonons}, Phys. Rev. Lett. {\bf 93}, 237401 (2004).

\bibitem{zpl1}M. Bayer and A. Forchel, {\it Temperature dependence of the exciton homogeneous linewidth in ${\mathrm{In}}_{0.60}{\mathrm{Ga}}_{0.40}\mathrm{As}/\mathrm{GaAs}$ self-assembled quantum dots}, Phys. Rev. B {\bf  65}, 041308 (2002)



\bibitem{10}
P. Machnikowski, {\em Change of Decoherence Scenario and Appearance of Localization due to Reservoir Anharmonicity},
Phys. Rev. Lett. {\bf 96}, 140405 (2006).

\bibitem{11} E. A. Muljarov and R. Zimmermann,
{\em Dephasing in Quantum Dots: Quadratic Coupling to Acoustic Phonons},
 Phys. Rev. Lett. {\bf 93}, 237401 (2004).
\bibitem{12} G. Ortner, D. R . Yakovlev, M. Bayer, S. Rudin, T. L. Reinecke,
S. Fafard, Z . Wasilewski, and A. Forchel,
{\em Temperature dependence of the zero-phonon linewidth in InAsGaAs quantum dots},  Phys. Rev. B {\bf 70}, 201301(R) (2004).
\bibitem{13} S. Rudin, T. L. Reinecke, and M . Bayer,
{\em Temperature dependence of optical linewidth in single InAs quantum dots},
 Phys. Rev. B 74, 161305(R)
(2006).
\bibitem{14} G. Lindwall, A. Wacker, C . Weber, and A. Knorr, 
{\em Zero-Phonon Linewidth and Phonon Satellites in the Optical Absorption of Nanowire-Based Quantum Dots},
Phys. Rev. Lett. {\bf 99}, 087401 (2007).


\bibitem{roy}C. Roy  and S. John,
{\it Microscopic theory of multiple-phonon-mediated dephasing and relaxation of quantum dots near a photonic band gap}, Phys. Rev. A {\bf 81}, 023817 (2010).

\bibitem{breuer} H.-P. Breuer, B. Kappler, and F. Petruccione, {\it Stochastic wave-function method for non-Markovian quantum master equations}, Phys. Rev. A., {\bf 59}, 1633 (1999)

\bibitem{QDParams}
We use the following phonon parameters for our
calculations: deformation potentials $D_e-D_h=6.5\,$eV,  mass density $\rho=5.667\,{\rm g\,cm^{-3}}$, longitudinal sound velocity $c_l=3800\,{\rm m\,s^{-1}}$. We find $\omega_{b}=1\,$meV and $\alpha_{p}/(2\pi)^2=0.06 \, {\rm ps}^{2}$ as typical numbers for InAs/GaAs quantum dots~\cite{JiaoJPC2008}, which we have also used to fit several
semiconductor  cavity-QED experiments~\cite{hughes2,roy_hughes}.

\bibitem{nazir3} D. P. S. McCutcheon, N. S. Dattani, E. M. Gauger, B. W. Lovett, and A. Nazir,
{\it A general approach to quantum dynamics using a variational master equation: Application to phonon-damped Rabi rotations in quantum dots}, Phys. Rev. B {\bf 84}, 081305(R) (2011)


\bibitem{makri1} N. Makri and D. E. Makarov,
{\it Tensor propagator for iterative quantum time evolution of reduced density matrices. I. Theory}, J.  Chem. Phys. {\bf 102} 4600 (1995)

\bibitem{makri2} N. Makri and D. E. Makarov,
{\it Tensor propagator for iterative quantum time evolution of reduced density matrices. II. Numerical methodology}, J. Chem. Phys. {\bf 102} 4611 (1995).


\bibitem{QOToolbox}
S M Tan, {\it A computational toolbox for quantum and atomic optics}, J. Opt. B: Quantum Semiclass. Opt. {\bf 1}, 424 (1999).




\bibitem{knorr2}K. J. Ahn,  J. F\"orstner,  and A. Knorr,
{\it Resonance fluorescence of semiconductor quantum dots: Signatures of the electron-phonon interaction}, Phys. Rev. B, {\bf 71}, 153309 (2005).


\end{thebibliography}
\end{document}